\documentclass[journal,twoside,web]{ieeecolor}
\usepackage{tmi}
\usepackage{cite}
\usepackage{amsmath,amssymb,amsfonts}
\usepackage{algorithmic}
\usepackage{graphicx}
\usepackage{textcomp}
\usepackage{algorithm}

\usepackage{adjustbox}
\usepackage{threeparttable}
\usepackage{booktabs}
\usepackage{multirow}
\usepackage{tabularx}
\usepackage{subfigure}

\usepackage{hyperref}
\hypersetup{hidelinks}

\def\BibTeX{{\rm B\kern-.05em{\sc i\kern-.025em b}\kern-.08em
    T\kern-.1667em\lower.7ex\hbox{E}\kern-.125emX}}
\begin{document}
\title{Disentangled Multi-modal Learning of Histology and Transcriptomics for Cancer Characterization}
\author{Yupei Zhang, Xiaofei Wang, Anran Liu, Lequan Yu, \IEEEmembership{Member, IEEE}, Chao Li
\thanks{This work was supported by the Guarantors of Brain. (Yupei Zhang and Xiaofei Wang contributed equally to this work.)}
\thanks{Y. Zhang and X. Wang are with Department of Clinical Neurosciences, University of Cambridge, UK (e-mail: yz931@cam.ac.uk).}
\thanks{A. Liu is with Department of Health Technology \& Informatics, The Hong Kong Polytechnic University, Hong Kong SAR, China (e-mail: anran.liu@connect.polyu.hk).} 
\thanks{L. Yu is with Department of Statistics and Actuarial Science, The University of Hong Kong, Hong Kong SAR, China (e-mail: lqyu@hku.hk).} 
\thanks{C. Li is with Department of Clinical Neurosciences and Department of Applied Mathematics and Theoretical Physics, University of Cambridge; School of Science and Engineering and School of Medicine, University of Dundee, UK (Corresponding author, e-mail: cl647@cam.ac.uk.)}
}

\maketitle

\begin{abstract}
Histopathology remains the gold standard for cancer diagnosis and prognosis. With the advent of transcriptome profiling, multi-modal learning combining transcriptomics with histology offers more comprehensive information. 
However, existing multi-modal approaches are challenged by intrinsic multi-modal heterogeneity, insufficient multi-scale integration, and reliance on paired data, restricting clinical applicability. 
To address these challenges, we propose a disentangled multi-modal framework with four contributions: 1) To mitigate multi-modal heterogeneity, we decompose WSIs and transcriptomes into tumor and microenvironment subspaces using a disentangled multi-modal fusion module, and introduce a confidence-guided gradient coordination strategy to balance subspace optimization. 2) To enhance multi-scale integration, we propose an inter-magnification gene-expression consistency strategy that aligns transcriptomic signals across WSI magnifications. 3) To reduce dependency on paired data, we propose a subspace knowledge distillation strategy enabling transcriptome-agnostic inference through a WSI-only student model. 4) To improve inference efficiency, we propose an informative token aggregation module that suppresses WSI redundancy while preserving subspace semantics. Extensive experiments on cancer diagnosis, prognosis, and survival prediction demonstrate our superiority over state-of-the-art methods across multiple settings. Code is available at \href{https://github.com/helenypzhang/Disentangled-Multimodal-Learning}{GitHub}.
\end{abstract}


\section{Introduction}
\label{sec:introduction}

\IEEEPARstart{H}{istopathology} remains the gold standard for cancer diagnosis and prognosis \cite{kather2019predicting}. However, conventional histopathological assessment is labor-intensive and subject to inter-observer variability, as it relies on individual expertise of pathologists. Computational pathology seeks to overcome these limitations by leveraging automated algorithms to analyze whole slide images (WSIs), enabling faster and more reproducible workflows. 
In particular, deep learning approaches, especially multiple instance learning (MIL)-based methods, have shown success in learning discriminative morphological representations for characterizing cancers \cite{chen2022scaling,ilse2018attention,shao2021transmil, zhang2022dtfd}. 
Early MIL utilizes mean or max pooling to aggregate instance-level features into a slide-level representation, ignoring structural and contextual relationships. Recent attention \cite{shao2021transmil, lu2021data} or graph-based \cite{li2024dynamic, chan2023histopathology, bontempo2023graph, qu2025multimodal} MIL methods explicitly model patch-patch correlations, thereby capturing spatial organization and tissue heterogeneity. To leverage the pyramidal structured WSI, Bontempo et al. \cite{bontempo2023graph} proposed a multi-scale graph-based MIL framework to model both inter- and intra-scale correlations.

In parallel, transcriptome profiling captures molecular-level cancer dynamics underlying tissue morphology. 
Transcriptomic modeling has increasingly incorporated biological structure, including gene-set aggregation \cite{chen2021multimodal, xu2023multimodal, qu2025memory, qu2025multimodal} and pathway-level representations \cite{jaume2024modeling, zhang2024prototypical, song2024multimodal}, to enhance interpretability in downstream tasks. These reflect a growing consensus that molecular signals are organized around functional programs rather than individual genes.
Histology and transcriptomics provide complementary perspectives on tumors, and integrating histological and transcriptomic features \cite{chen2021multimodal,zhou2023cross,song2024multimodal, xiong2024mome, jaume2024modeling} thus holds promise for more comprehensive and precise cancer characterization. 
Representative multi-modal methods employ attention-based fusion, co-attention transformers \cite{chen2021multimodal}, optimal transport \cite{xu2023multimodal}, as well as graph- \cite{qu2025memory} or hypergraph-based \cite{qu2025multimodal} formulations to model cross-modal interactions.

Despite the potential, existing multi-modal learning methods face challenges in multi-modal modeling, integration, and applicability.
\textbf{1) Modeling tumor heterogeneity across modalities:}
Tumor ecosystems comprise diverse cellular populations, including both tumor and microenvironment components \cite{hu2023deciphering}, manifesting both morphological and transcriptomic features. While WSIs and transcriptomes offer comprehensive information, it remains challenging to model their complex associations. Current methods \cite{chen2020pathomic} often fail to disentangle the contributions of cellular sources from the tumor and microenvironment. This neglect of biological semantics limits interpretability and may degrade predictive performance. 

\textbf{2) Integrating transcriptome with multi-scale WSI:}
WSIs inherently encode diagnostically relevant information at multiple spatial scales.
Lower magnifications capture global tissue architecture, while higher magnifications encode fine-grained cellular morphology~\cite{bontempo2023graph}.
Relying on a single magnification therefore provides an incomplete characterization of tissue heterogeneity.
Transcriptomic measurements reflect molecular underpinnings across heterogeneous tissue regions and consequently exhibit biologically meaningful correlations across WSI scales. Aligning molecular information to WSI at a fixed magnification may introduce bias.
Therefore, it is necessary to explicitly model multi-scale WSI and transcriptome correlation.
However, existing multi-modal methods typically process WSIs at a single scale or naively aggregate multi-scale features with transcriptome without biological constraints, leading to insufficient multi-scale fusion.

\textbf{3) Reducing reliance on transcriptome during inference:} 
In real-world settings, transcriptome profiling is often unavailable due to cost, tissue constraints, or turnaround time. Most current multi-modal models, however, assume availability of paired WSI-transcriptome \cite{xing2022discrepancy}, limiting their translational viability. It is essential to develop models that are transcriptome-agnostic during inference. 
Yet, effectively transferring transcriptome-informed supervision to WSI-only inference remains an open challenge.

\textbf{4) Reducing redundancy in WSI-based inference:}
The gigapixel WSIs contain rich morphological information, yet introduce redundant or non-discriminative features, obscuring diagnostically important but spatially sparse features \cite{zheng2024deep}. 
Traditional MIL approaches apply mean or max pooling on patch embeddings  \cite{oner2023distribution,campanella2019clinical}, which do not address the redundancy.
Recent attention-based pooling, while providing adaptive weighting capabilities~\cite{shao2021transmil}, is constrained by rigid receptive fields in capturing sparsely distributed but critical variations.
Effectively identifying and reducing redundancy remains challenging for WSI representation.

To address these challenges, we propose a biologically inspired, two-stage framework that learns complementary tumor and microenvironment representations across WSIs and transcriptomics, while enabling WSI-only inference.
\textbf{Firstly}, to capture tumor heterogeneity across modalities, we explicitly decompose transcriptome into two subspaces: tumor and tumor microenvironment (TME), reflecting distinct yet complementary components \cite{hu2023deciphering}. 
Within each subspace, we propose a Disentangled Multi-modal Selective Fusion (DMSF) module to identify and integrate informative multi-modal features. To balance inter-subspace optimization, we introduce a Confidence-guided Gradient Coordination (CGC) strategy, adjusting subspace gradients based on predicted reliability. 
\textbf{Secondly}, to align transcriptomes with multi-scale WSI features, we propose the Inter-magnification Gene-expression Consistency (IGC) strategy, which encourages consistency in transcriptome attention across WSI magnifications,  reflecting biological coherence of gene-expression signals across tissue scales. A Diagonal Element Variance (DEV) Loss enforces this consistency, enhancing robust multi-scale integration.
\textbf{Thirdly}, to ensure applicability when transcriptome is unavailable in inference, we propose a Subspace Knowledge Distillation (SKD) strategy. During training, a teacher model is exposed to both WSIs and transcriptome, and transfers subspace knowledge to a WSI-only student model for inference.
\textbf{Lastly}, to reduce WSI redundancy and enable efficient inference, we introduce an Informative Token Aggregation (ITA) module. Instead of applying attention across all patches, ITA uses a deformable attention to encourage models to focus on diagnostically critical patches. 
Extensive experiments on both cancer diagnosis and prognosis tasks in three public datasets demonstrate that our method outperforms competing methods. 

This work substantially extends our previous conference paper~\cite{zhang2024knowledge}. \textbf{At the methodological level}, we 1) improved the clinical feasibility of multi-modal learning under missing transcriptomic inputs by introducing a student model trained via SKD strategy, 2) enhanced multi-scale feature integration through an IGC strategy, and 3) refined multi-modal fusion with a transcriptome-selective module. \textbf{At the experimental level}, we 1) extended the evaluation by including additional state-of-the-art (SOTA) baselines, considering uni-modal, missing-modality, and multi-modal settings, 2) added survival analysis with external validation, and 3) improved model performance across diagnosis, grading, and prognosis tasks. \textbf{At the presentation level}, we 1) added the related work section, 2) introduced additional preliminary findings that motivate key design choices, and 3) provided more in-depth interpretability studies. Together, these additions constituted a substantial extension beyond the conference version.

\section{Related Work}
\subsection{WSI-based Precision Oncology}

WSIs provide rich morphological information critical for precision oncology. Deep learning methods have been developed based on WSIs for prediction tasks. 
Earlier studies focused on region-of-interests (ROI) to localize diagnostically important regions \cite{kather2019predicting,chen2020pathomic,xing2022discrepancy,pan2024focus}. However, manually delineating ROI was labor-intensive and relied on experts. To address this, MIL-based approaches \cite{ilse2018attention, shao2021transmil, chen2022scaling, coudray2018classification} learned slide-level embeddings by aggregating patch-level features. 
Nevertheless, the gigapixel size of WSIs introduced redundancy, hindering efficient representation learning. 
To alleviate this challenge, we propose an ITA module, which identifies and groups informative tokens into representative prototypes for efficient WSI-based representation. Notably, as modern cancer diagnostics increasingly combine histology and molecular markers, there is an urgent need for effective multi-modal frameworks integrating WSI and transcriptome.
\begin{figure*}[!t]
\centering
\includegraphics[scale=.34]{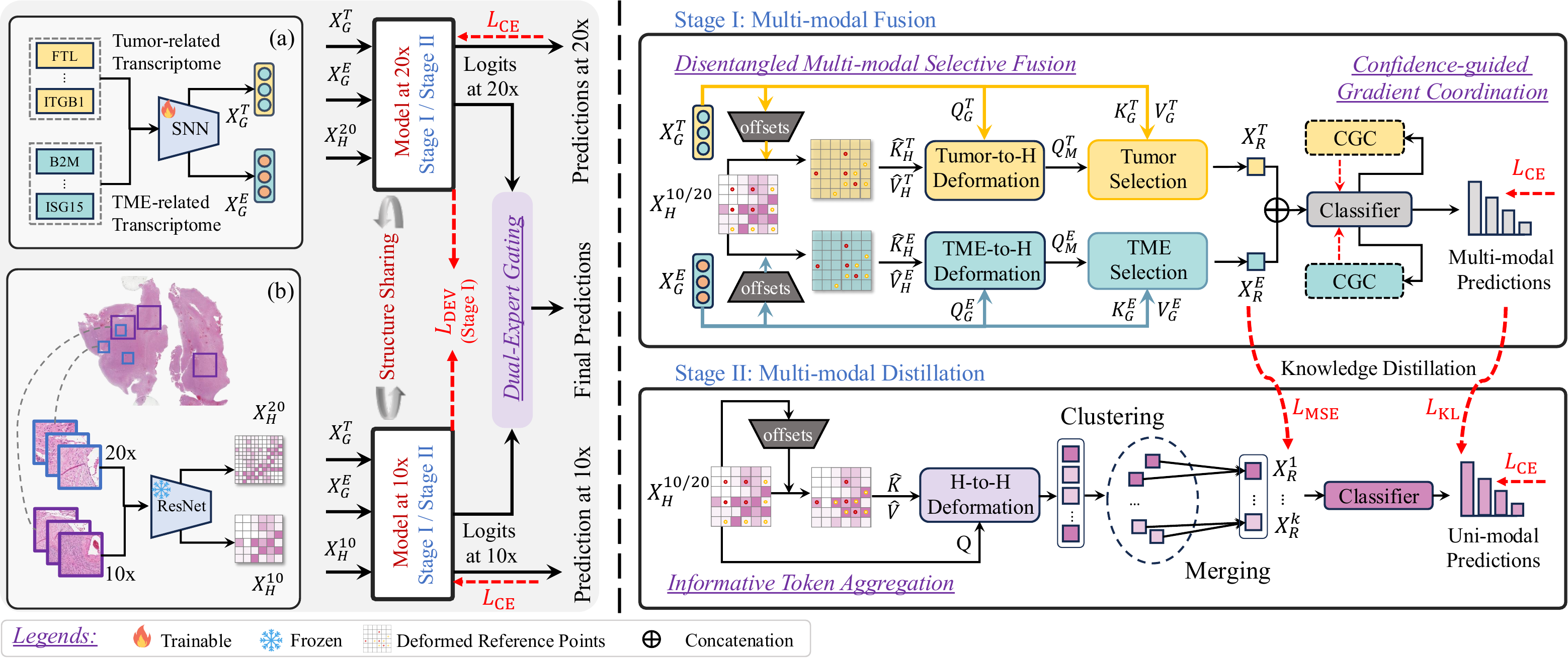}
\caption{Framework overview. \textbf{Left:} Multi-modal inputs (with (a) disentangled transcriptomic profiles and (b) multi-scale WSI embeddings at $10\times$ and $20\times$ magnification), and Multi-scale architecture (with the DEV loss applied across scales in Stage~I). \textbf{Right:} Two-stage framework, where Stage~I learns subspace-aware multi-modal representations and Stage~II performs multi-modal distillation. Only WSIs are required at inference.}
\label{fig:framework}
\end{figure*} 

\subsection{Multi-modal Learning with WSI and Transcriptomics}
Integrating multi-modal data promises to promote precision oncology~\cite{chen2021multimodal, zhou2023cross}. Earlier efforts primarily focused on single modalities.
For WSI modeling, MIL was commonly used to derive slide-level features, while for transcriptome, recent approaches ~\cite{zhou2023cross} developed biologically structured representations, such as grouping genes into broad functional families~\cite{chen2021multimodal}, or pathways~\cite{song2024multimodal}.
To bridge modalities, earlier methods fused features via concatenation or addition. However, these approaches are limited by substantial gaps between WSIs and transcriptomes. Recent approaches~\cite{chen2021multimodal, zhou2023cross} sought to alleviate this gap via cross-modal alignment mechanisms. For instance, recent efforts~\cite{song2024multimodal} introduced multi-modal prototypes and OT-based cross-alignment to improve integration.
However, these methods overlooked shared semantic structures that underpin both modalities. 
In this study, we address this gap by explicitly modeling tumor and TME subspaces~\cite{hu2023deciphering, yuan2024beyond}, which serve as shared semantic anchors across modalities. This design captures diagnostically specific features and supports more coherent multi-modal representation learning.
Further, we propose the IGC strategy to enhance multi-scale integration (integrating transcriptome and multi-scale WSI). 

\subsection{Multi-modal Learning with Missing-Modality}
Despite success, most multi-modal methods face challenges in translation due to their reliance on paired multi-modal data at inference, particularly given the limited availability of transcriptome in clinical practice. To address this, Xing et al. \cite{xing2022discrepancy} proposed a distillation framework to transfer knowledge from a multi-modal teacher to a uni-modal student for glioma grading. Pan et al. \cite{pan2024focus} proposed gene-mutation guided to encourage the model to focus on discriminative ROI.
Wang et al. \cite{wang2023multi} proposed a multi-task framework to predict molecular markers and glioma classification from WSIs, requiring WSIs only during inference. 
More recently, DDM-net \cite{qiu2024dual} proposed an imputation method to handle missing genomic or pathology data, although such approaches are hindered by large multi-modal heterogeneity. 
Similarly, LD-CVAE \cite{zhou2025robust} is a multi-modal survival prediction framework that reconstructs genomic representations from WSIs via a conditional VAE and integrates them with co-attention–based instance selection to enable robust WSI-driven survival prediction under missing genomic data.
G-HANet \cite{wang2025histo} proposed a genome-informed hyper-attention framework that distills histo-genomic associations by reconstructing genomic representations from WSIs during training, and leverages the distilled knowledge to enhance WSI-based hyper-attention modeling for survival prediction at inference time.
In contrast, we distilled subspace-specific representations, capturing tumor and TME semantics across both WSIs and transcriptomes, into a WSI-only student, enabling biologically meaningful knowledge distillation.
\begin{figure*}[!t]
\centering
\includegraphics[scale=.42]{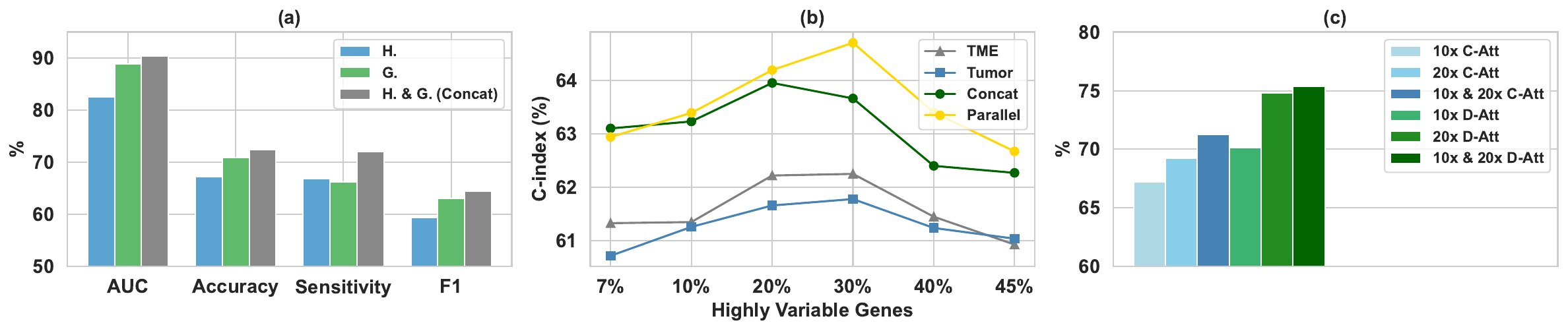} 
\caption{Findings for model design. (a) Performance on cancer diagnosis with WSI-only, transcriptomic-only, and WSI-transcriptome integration by concatenation. (b) C-index of different input forms of tumor and TME-related genes. (c) Accuracy with different scales WSIs as input in cancer grading with cross-attention (C-Att) or deformable attention (D-Att) attention mechanisms.} \label{finding}
\end{figure*}

\section{Methodology}
\subsection{Preliminary Findings for Model Design}
To guide the development of our model, we conduct empirical analyses using the TCGA GBM-LGG dataset to examine cross-modal and intra-modal relationships, yielding three key findings that motivated our design.

\textbf{1. Multi-modal integration outperforms single-modality}.
\label{method:findings1} 
To assess the benefit of multi-modal integration, we evaluate a transformer-based model with three input configurations: transcriptomics-only, WSI-only, and a concatenated combination of both. As shown in Fig.~\ref{finding} (a), the multi-modal input significantly outperforms either single modality in glioma diagnosis, highlighting the need for multi-modal integration. This observation is consistent with prior work \cite{chen2020pathomic} and motivates our proposed distillation framework: the multi-modal teacher could benefit the WSI-only student that remains clinically deployable.

\textbf{2. Disentangled modeling of tumor and TME-related genes improves prognostics}.
\label{method:findings2}
Guided by biological priors, we decompose transcriptomic data into tumor and TME-related gene sets \cite{hu2023deciphering} and test three input strategies using a Self-Normalizing Network (SNN) \cite{klambauer2017self}: 
\textit{1) separate}: input each gene set individually; \textit{2) concat}: concatenate both gene sets prior to input; \textit{3) parallel}: process each set in parallel using two separate SNNs and fuse their representations via a multilayer perceptron (MLP). As shown in Fig. \ref{finding} (b), the \textit{parallel} strategy shows the best performance in survival prediction, especially with $30\%$ highly variable genes, motivating our design of disentangled and parallel processing of genes in tumor and TME subspaces in Section \ref{method:DMSF}.

\textbf{3. Focusing on informative multi-scale patches enhances performance}.
\label{method:findings3} 
To enhance focus on informative regions on WSIs, we compare standard cross-attention and deformable attention \cite{xia2022vision} on multi-scale WSI features. As shown in Fig. \ref{finding} (c), deformable attention yields superior performance, indicating its effectiveness in selectively attending to critical visual patterns. These insights motivate our use of dynamic deformable attention (deformation layer in Section \ref{method:DMSF}) and multi-scale inputs (Section \ref{method:IGC}) to capture both coarse and fine-grained histological structures.

\subsection{Model Overview}
As illustrated in Fig.~\ref{fig:framework}, we propose a two-stage framework for integrating WSIs and transcriptomes, supporting both multi-modal learning and efficient WSI-only inference. Both stages operate at $10\times$ and $20\times$ magnifications and share the same architecture.
\textbf{Stage I} (Section \ref{method:StageI}) performs multi-modal fusion, taking WSI features ($X_H^{10}$, $X_H^{20}$) and transcriptomes embeddings split into tumor ($X_G^T$) and TME ($X_G^E$) gene subsets. Within each subspace, the DMSF module (Section \ref{method:DMSF}) fuses modalities via selective attention and generates subspace representations ($X_R^T$, $X_R^E$). To balance subspaces optimization, a CGC strategy (Section \ref{method:CGC}) adjusts gradients based on subspace prediction confidence; and the IGC strategy (Section \ref{method:IGC}) further enforces transcriptome-guided attention consistency to enhance multi-scale consistency.
\textbf{Stage II} (Section \ref{method:StageII}) enhances clinical applicability using WSI alone. A WSI-only model is first warmed up and then refined via SKD (Section \ref{method:SKD}), transferring subspace knowledge from a multi-modal teacher. Unlike standard distillation, SKD explicitly preserves subspace semantics to retain biological interpretability. To reduce redundancy, the student employs an ITA module (Section \ref{method:ITA}), which clusters and merges patch tokens into subspace-aware morphological prototypes ($X_R^{K}$, where $K \in \{1,k\}$). 
Final predictions from both magnifications are combined via a dual-expert gating module, which fuses the logits from the $10\times$ and $20\times$ branches in logit space using a confidence-based rule by default, while each magnification branch is independently supervised during training to ensure stable optimization.
To this end, the student model enables transcriptome-informed learning, allowing for interpretable and efficient inference using WSI alone.

\subsection{Stage I: Multi-modal Fusion}
\label{method:StageI}
To tackle challenges of multi-modal heterogeneity and multi-scale integration, Stage I performs multi-modal modeling and fusion by disentangling biologically grounded subspaces and aligning representations across scales. 
\subsubsection{Disentangled Multi-modal Selective Fusion} \label{method:DMSF}
Motivated by biological prior of tumor and TME compartmentalization \cite{burgos2019tumour, yuan2024beyond, hu2023deciphering} in both histology and transcriptomics, and the parallel gene inputs in finding 2 (Section \ref{method:findings2}.2), the DMSF module introduces two branches to explicitly model subspace-specific multi-modal representations, capturing tumor-related (T subspace) and TME-related (E subspace) characteristics, respectively.
Within each subspace, a two-step multi-modal selective fusion module is implemented to selectively integrate informative features from histology and transcriptomics. According to finding 3 (Section \ref{method:findings3}.3), we adopt the deformable attention layer for WSI feature representation.
For example, T subspace includes: 1) \textit{A Tumor-to-H Deformation layer} that identifies informative WSI features (H: histology) guided by transcriptomics and 2) \textit{A Tumor Selection layer} that selects task-relevant transcriptomic features. 
The following description focuses on T subspace at $10\times$ magnification, with E subspace handled similarly. For clarity, we use $X_H^T$ to denote input WSI features instead of $X_H^{10;T}$.

\textbf{First}, the \textit{Tumor-to-H Deformation layer} uses transcriptome features $X_{G}^T$ to generate spatial offsets $\Delta p^T$ via a learnable module $\Psi$, consisting of two convolution layers and a scaler, which guides deformable sampling over WSI features $X_H^T$. Given $X_{G}^T \in \mathbb{R}^{h \times w \times c}$, the initial reference points $p^T \in \mathbb{R}^{h_G \times w_G \times 2}$ form a uniform grid. 
The deformed histology features are then sampled: $\hat{X}_H^T=F(X_H^T; \ {\rm norm}(p^T+\Delta p^T))$, $\Delta p^T = \Psi(X_{G}^T)$, where $F$ is a bilinear interpolation sampling function.
The query, deformed key and value in the multi-head transcriptome to histology deformable attention are:
\begin{equation} \label{e:}
    Q^T_G = X_{G}^T W_{Q}^T, \ \hat{K}_H^T = \hat{X}_H^T W_{K}^T, \ \hat{V}_H^T = \hat{X}_H^T W_{V}^T,
\end{equation}
where $W_Q^T, W_K^T, W_V^T$ are corresponding projection networks. Moreover, the output of one attention head is:
\begin{equation} \label{e:}
    Z_M^{I;T} = {\rm softmax}(Q^{(I);T}\hat{K}^{(I);T \top} / \sqrt{d})\hat{V}^{(I);T},
\end{equation}
where the attention head index is denoted as $I$, with $I \in \{1, 2, \ldots, i\}$, and $M$ represented multi-modal. The multi-modal outputs are obtained by:
\begin{equation} \label{e:}
    Z_M^T = {\rm concat}(Z_M^{1;T}, ..., Z_M^{i;T}) W_M^T,
\end{equation}
where $W_M^T$ is the projection network. Accordingly, this process integrates spatially deformed WSI features guided by transcriptomic context.

\textbf{Second}, the \textit{Tumor Selection layer} enables selective attention to transcriptome features by multi-modal query $Q_M^T$, thereby further refining the fused representation by attending to task-relevant transcriptomic features. The query $Q_M^T$ is derived from $Z_M^T W_Q^T$ ($W_Q^T$ is the projection network), while transcriptome features $X_G^T$ are similarly projected using $W_K^T$ and $W_V^T$ to obtain keys $K_G^T$ and values $V_G^T$.
\begin{equation} \label{e:}
    Z_O^{I;T} = {\rm softmax}(Q_M^{(I);T}K_G^{(I);T \top} / \sqrt{d})V_G^{(I);T},
\end{equation}
\begin{equation} \label{e:}
    Z_O^T = {\rm concat}(Z_O^{1;T}, ..., Z_O^{i;T}) W_O^T,
\end{equation}
where $O$ represents the output. 
Together, these two layers promote fine-grained, bidirectional integration of morphological and molecular features within each biological subspace, enhancing multi-modal fusion.

\subsubsection{Confidence-guided Gradient Coordination} \label{method:CGC}
\begin{figure}[!t]
\centering
\includegraphics[scale=.35]{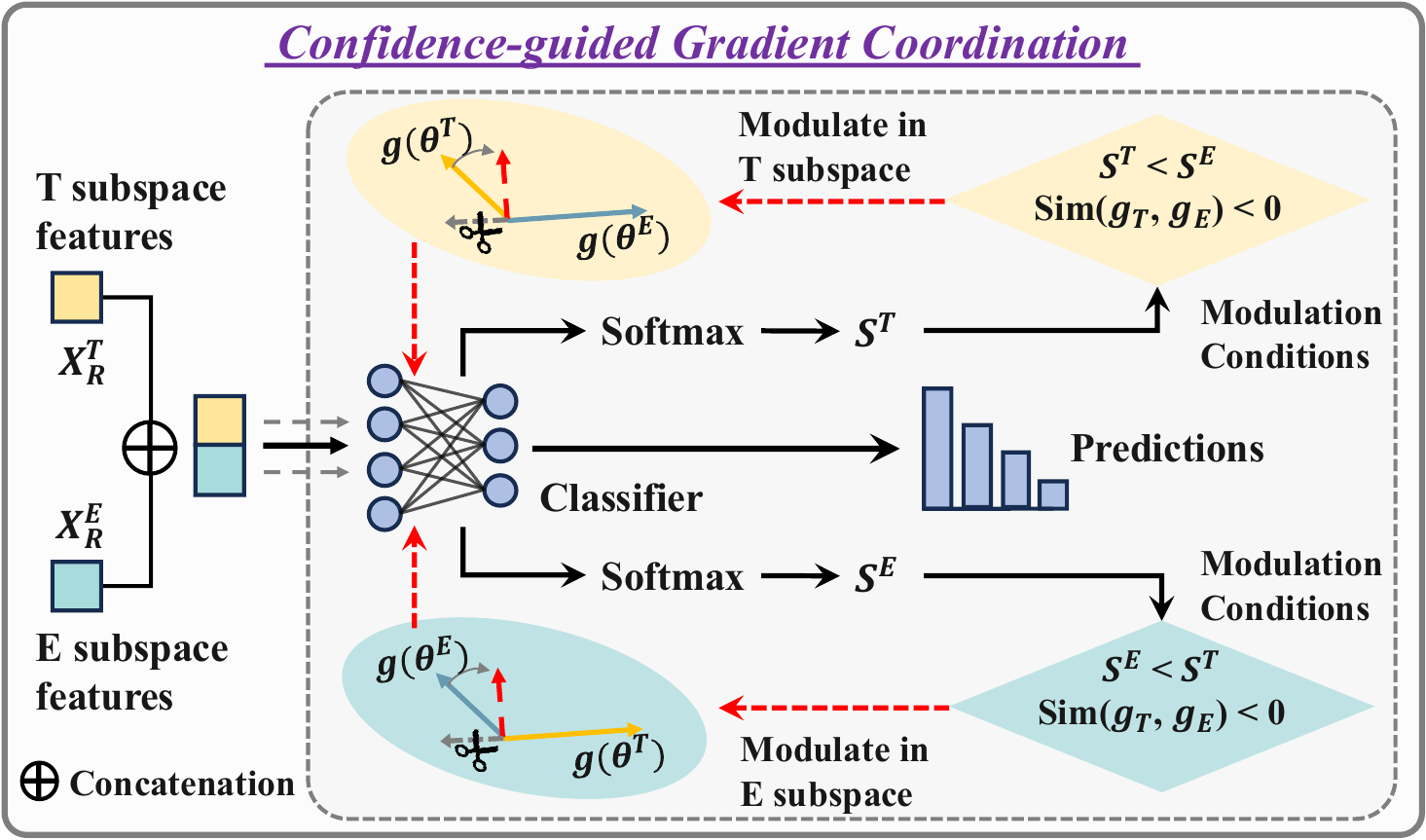} 
\caption{The Confidence-guided Gradient Coordination strategy. The confidence scores $S^T$ and $S^E$ are calculated with softmax after subspace logits. The less confident gradient is projected onto the orthogonal complement of the more confident one.} \label{fig:CGC}
\end{figure}

Despite DMSF disentangles tumor and TME subspaces, joint optimization of the subspaces can suffer from gradient conflicts during training, impeding global optimization. To address this, we propose a CGC strategy that resolves conflicting gradients based on predictive confidence. As shown in Fig. \ref{fig:CGC}, the cosine similarity between the subspace gradients $g(\theta^T)$ and $g(\theta^E)$ is calculated as ${\rm cosine}(g(\theta^T), g(\theta^E))$, where a value less than zero indicates gradient conflict. 
To assess reliability, we compute confidence scores $S$, defined as the predictive probability of the given true label, where $S^T = {\rm softmax}(\mathcal{D}(X^T_R))[l]$, $S^E = {\rm softmax}(\mathcal{D}(X^E_R))[l]$, and $\mathcal{D}$ is the downstream classifier. 
Summing over a mini-batch,  $\sum S^T$ and $\sum S^E$ represent the batch-level confidence on the $l$-$th$ label, respectively.

If a conflict occurs, the less confident gradient is projected onto the orthogonal complement of the more confident one:
\begin{equation} \label{e:cgc}
    \begin{cases}
        \ \tilde{g}(\theta^T) \ = \ \gamma (g(\theta^T), \ g(\theta^E)), & {\sum S^T < \sum S^E}, \\
        \ \tilde{g}(\theta^E) \ = \ \gamma (g(\theta^E), \ g(\theta^T)), & {\sum S^E < \sum S^T},
    \end{cases}
\end{equation}
where $\gamma(\vec{x}_1, \vec{x}_2)$ represents the projection of the vector $\vec{x}_1$ onto the orthogonal complement to the vector $\vec{x}_2$.
This dynamic adjustment ensures smooth and confidence-aware coordination of subspace learning.

\subsubsection{Inter-magnification Gene-expression Consistency}
\label{method:IGC}
\begin{figure}[!t]
\centering
\includegraphics[scale=.323]{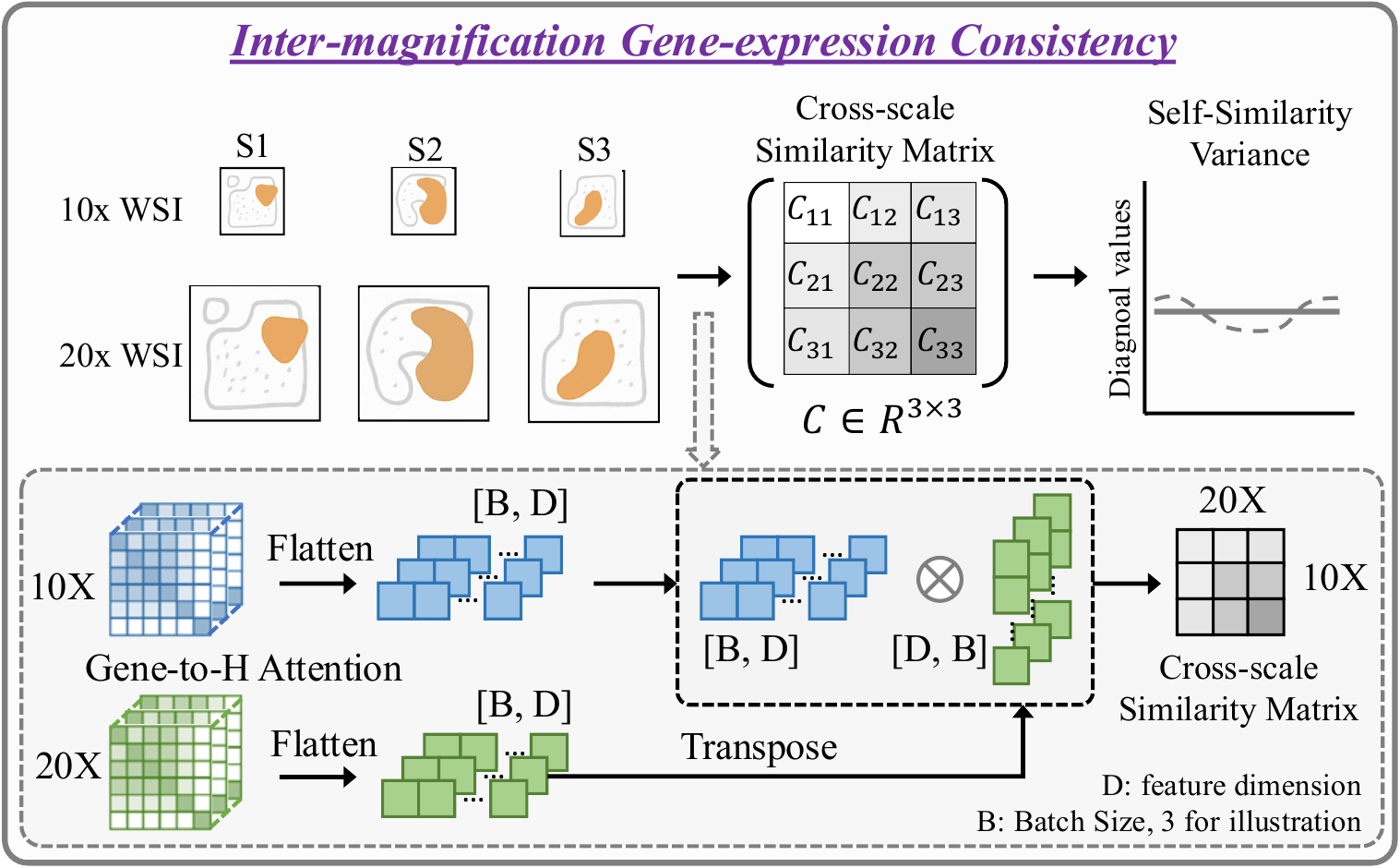} 
\caption{The Inter-magnification Gene-Expression Consistency Strategy. A cross-scale similarity matrix is utilized to measure the inter-magnification consistency. The sample-wise inter-magnification consistency is constrained by a Diagonal Element Variance loss.} \label{fig:IGC}
\end{figure}

While DMSF and CGC align modalities within each subspace, they do not enforce consistency across magnification levels. According to finding 3 (Section \ref{method:findings1}.3), we utilize the multi-magnification WSI.
As transcriptomics activities are consistently reflected by WSI across magnifications, we propose the IGC module to encourage biologically meaningful integration across scales. As shown in Fig. \ref{fig:IGC}, given the Tumor-to-H attentions or TME-to-H attentions on multi-scale WSI features, we first flatten them to obtain tumor-wise multi-scale weights ($\mathbf{A}_{G^{T};H^{10}} \in \mathbb{R}^{B\times D}$, $\mathbf{A}_{G^{T};H^{20}} \in \mathbb{R}^{B\times D}$), and the TME-wise multi-scale weights ($\mathbf{A}_{G^{E};H^{10}} \in \mathbb{R}^{B\times D}$, $\mathbf{A}_{G^{E};H^{20}} \in \mathbb{R}^{B\times D}$), where $B$ is the number of samples, $D$ is the dimension after flatten. 
To measure intra-sample consistency, we compute the cross-scale similarity metric $C \in \mathbb{R}^{B\times B}$: $\mathbf{A}_{G^{T};H^{10}} \cdot (\mathbf{A}_{G^{T};H^{20}})^{\top}$ or $\mathbf{A}_{G^{E};H^{10}} \cdot (\mathbf{A}_{G^{E};H^{20}})^{\top}$. Of note, the diagonal elements of $C$ reflect similarity between $10\times$ and $20\times$ magnifications of a specific gene set. 
Finally, we introduce a Diagonal Element Variance (DEV) loss:
\begin{equation}
\mathcal{L}_{\text{DEV}} = \lambda \cdot \frac{1}{n} \sum_{i=1}^n \left( C_{ii} - \frac{1}{n} \sum_{j=1}^n C_{jj} \right)^2
\end{equation}
This soft regularization penalizes deviations from the average intra-sample consistency, grounded in the hypothesis that each gene group should consistently attend to multi-scale WSI features, encouraging robust multi-scale alignment.

\subsection{Stage II: Multi-modal Distillation}
\label{method:StageII}
Stage I enables multi-modal learning across tumor and TME subspaces. According to the multi-modal potential in finding 1 (Section \ref{method:findings1}.1), in Stage II, we focus on improving clinical applicability by introducing a WSI-based student model (Section~\ref{method:ITA}) with subspace knowledge distilled. This is achieved through the ITA module (Section~\ref{method:ITA}) and SKD strategy (Section~\ref{method:SKD}).
\subsubsection{Informative Token Aggregation} \label{method:ITA}
The student model consists of ITA module, which identifies and aggregates representative WSI regions into morphological prototypes. As depicted in Fig. \ref{fig:framework}, ITA contains two stages: \textit{Informative Token Learning} (i.e., H-to-H Deformation) and \textit{Morphological Prototype Aggregating} (i.e., Clustering and Merging).

The \textit{Informative Token Learning} encourages the model to focus on the spatially informative patches through a deformable attention layer. Taking $10\times$ magnification as an example, the offsets are generated by the offsets generation network $\Psi$, with the guidance of WSI features $X_{H}^{10} \in \mathbb{R}^{h \times w \times c}$ (simplified as $X_{H}$). Then, the deformed features $\hat{X}_H$ are sampled via $F(X_H; \ {\rm norm}(p+\Delta p))$, with initial and deformed reference points $p \in \mathbb{R}^{h_G \times w_G \times 2}$, $\Delta p = \Psi(X_{H})$, and $F$ is a bilinear interpolation function. 
With query $Q = X_{H} W_{Q}$, deformed key $\hat{K} = \hat{X}_H W_{K}$, deformed value $\hat{V} = \hat{X}_H W_{V}$, the H-to-H Deformation is implemented with a deformable attention layer, and the output $Z^{I}$ of one attention head and the final output $Z$ are denoted as:
\begin{equation} \label{e:}
    Z^{I} = {\rm softmax}(Q^{(I)}\hat{K}^{(I) \top} / \sqrt{d})\hat{V}^{(I)},
\end{equation}
\begin{equation} \label{e:}
    Z = {\rm concat}(Z^{1}, ..., Z^{M}) W_O,
\end{equation}
where $I \in \{1,2,\ldots,i\}$ indexes the attention heads, and $W_Q$, $W_K$, $W_V$, and $W_O$ are the corresponding projection layers. This H-to-H deformation guides the model to focus on informative WSI regions, reducing redundancy.

In the \textit{Morphological Prototype Aggregating} module, we group informative features $Z$ into $K$ clusters, and all tokens in the $k$-th cluster are merged into a representative token $X_R^k$.
Specifically, we perform a density peak clustering with K-nearest neighbors (DPC-KNN \cite{du2016study}), based on feature similarity. Given $Z$, the token distance is computed as $D_{i,j} = \| Z_i - Z_j \|_2$, and the local density $\rho_i$ of the $i$-th token $Z_i$ is calculated as:
\begin{equation} \label{e:localdensity}
    \rho_i = \exp(- \frac{1}{k} \sum_{Z_j \in {\text{KNN}}(Z_i)} D_{i,j}^2), 
\end{equation}
where ${\rm KNN}(Z_i)$ denotes the $K$ nearest neighbors of the $i$-th token. For the token with the highest local density, the relative distance $\xi$ is defined as the maximum distance to all other tokens. For tokens with lower local density, $\xi$ is defined as the minimum distance to any token with higher local density. Detailed process for obtaining $\xi$ of each token is as:
\begin{equation} 
\label{e:relativedistance}
    \xi_i = 
    \begin{cases}
        \max_j D_{i,j}^2, & \text{if } \rho_i \text{ is maximum} \\
        \min_{j:\rho_j > \rho_i} D_{i,j}^2, & \text{if } \exists \, \rho_j > \rho_i 
    \end{cases}
\end{equation}
The cluster centers of tokens are selected with a higher local density and a larger relative distance from other tokens with higher densities. 
The representative score $s_i$ is defined as $\rho_i \times \xi_i$, representing the confidence of token $Z_i$ to be chosen as one of the cluster centers. The top-K highest tokens are selected as cluster centers. Inspired by previous work \cite{rao2021dynamicvit}, we predict the significance score $\omega$ of each token in the same cluster. The merged representation token for $k$-$th$ cluster $K_k$ is:
\begin{equation} \label{e:score}
     X_{R}^k= \frac{\Sigma_{i \in K_k} \omega_i Z_i }{\Sigma_{i \in K_k} \omega_i}, 
\end{equation}
where $i$ represents tokens belonging to the $K_k$. This design enables subspace knowledge learning from teacher models with the following distillation strategy. 

\subsubsection{Subspace Knowledge Distillation} \label{method:SKD}

To distill subspace knowledge from the multi-modal teacher to the WSI-only student, we use a hybrid distillation strategy comprising prediction-level and representation-level supervision. 
At the prediction level, we apply temperature-scaled softmax to the teacher’s logits \( z_i \) using temperature \( \tau \):
\begin{equation}
P_{\text{soft}}(i) = \frac{e^{z_i/\tau}}{\sum_{j} e^{z_j/\tau}},
\end{equation}
and minimize the Kullback–Leibler (KL) divergence \cite{hinton2015distilling} between the softened teacher outputs and the student predictions:
\begin{equation}
\mathcal{L}_{\text{KL}} = \sum_{i} P_{\text{soft}}(i) \log \frac{P_{\text{soft}}(i)}{P_{\text{student}}(i)}.
\end{equation}

At the representation level, we encourage the student to learn from teacher’s concatenated subspace features, \( \hat{X}_R = [X_R^T; X_R^E] \in \mathbb{R}^{B \times 256} \), using Mean Squared Error (MSE) loss
$\mathcal{L}_{\text{MSE}} = \| \hat{X}_R - X_R \|^2$.
This dual-objective distillation framework enables the student to learn both the final prediction space and subspace-specific semantics, addressing missing transcriptomes during inference.

\subsection{Training Objectives of Downstream Tasks}
\subsubsection{Stage I}
Task-specific loss functions are devised for downstream tasks. For diagnosis and grading, we adopt cross-entropy loss, and the overall training objective is defined as:
\begin{align}
    \mathcal{L}_{\text{diag}} &= \mathcal{L}_{\text{CE}}(\mathcal{D}(Z^T, Z^E; \ \theta_{\text{diag}}), \ Y_{\text{diag}}) + \mathcal{L}_{\text{DEV}}, \label{e:loss_diag}\\
    \mathcal{L}_{\text{grad}} &= \mathcal{L}_{\text{CE}}(\mathcal{D}(Z^T, Z^E; \ \theta_{\text{grad}}), \ Y_{\text{grad}}) + \mathcal{L}_{\text{DEV}}, \label{e:loss_grad}
\end{align}
where $\mathcal{L}_{\text{CE}}$ denotes the cross-entropy loss, $\mathcal{D}$ denotes the classifier, $\theta_{\text{diag}}$ and $\theta_{\text{grad}}$ correspond to diagnosis and grading parameters, while $Y_{\text{diag}}$ and $Y_{\text{grad}}$ are ground-truth labels.
For prognosis, we adopt the negative log-likelihood (NLL) survival loss \cite{zhou2023cross}, denoted as $\mathcal{L}_{\text{NLL}}$, as the task-specific objective. The final loss is formulated as:
\begin{equation}
    \mathcal{L}_{\text{surv}} = \mathcal{L}_{\text{NLL}} + \mathcal{L}_{\text{DEV}}, \label{e:loss_surv}
\end{equation}

\subsubsection{Stage II} In the second stage, we first warm up the uni-modal student with task-specific loss $\mathcal{L}_{\text{Task}}$ and then distill the pre-trained multi-modal subspace knowledge to the student with the following objectives:
\begin{equation}
    \mathcal{L}_{\text{MM-Distill}} = \mathcal{L}_{\text{Task}} + \mathcal{L}_{\text{MSE}} + \mathcal{L}_{\text{KL}}, \label{e:loss_MMDistill}
\end{equation}
where $\mathcal{L}_{\text{Task}}$ represents cross-entropy loss for diagnosis or grading, and NLL for prognosis.

\section{Experiments \& Results}
\subsection{Experimental Settings}
\subsubsection{Setup and Evaluation}
We evaluated our model on three tasks: glioma diagnosis, grading, and survival prediction on a meta-dataset and external validation.
Each task was assessed under three settings: \textbf{uni-modal} (WSI-only training and inference) to evaluate student model, \textbf{missing-modality} (multi-modal training, WSI-only inference) to evaluate the effectiveness of distillation, and \textbf{multi-modal} (WSI + transcriptome training and inference) to evaluate the multi-modal teacher. Transcriptomics-only training and inference are also conducted to benchmark its standalone performance.

For glioma diagnosis, we followed the 2021 WHO criteria with four labels: glioblastoma, oligodendroglioma, astrocytoma (grade 4), and low-grade astrocytoma, and three grades for the grading task: grade II,  III, and  IV.
Evaluation metrics included AUC, Accuracy, Sensitivity, Specificity, and F1-score.
For survival prediction, we employed a discrete-time survival model that outputs hazard probabilities across time intervals, following \cite{chen2023transformer}. Performance was evaluated using the concordance index (C-Index). Zero-shot generalization was evaluated on an independent dataset for survival prediction.

\subsubsection{Datasets}
We included three public datasets: TCGA GBM-LGG  \cite{tomczak2015review}, IvyGAP \cite{puchalski2018anatomic}, and CPTAC \cite{li2023proteogenomic}. 
IvyGAP includes only GBM cases and has a limited sample size, restricting its use for diagnostic tasks.
We therefore merged TCGA GBM-LGG and IvyGAP into a meta-dataset for internal validation, comprising 2,387 paired WSIs and transcriptome profiles from 668 cases.
CPTAC served as an external cohort for zero-shot validation.
\begin{table*}[!t]
 \scriptsize
    \caption{Comparison with SOTA methods on the Diagnosis task (3-fold validation). H./G. represents histopathology/genomics modalities. In our approach, the student, distillation, and teacher models are denoted as Stu, Dst, and Tch, respectively. Best and second results are in \textbf{bold} and \underline{underline}.} \label{table_diagnosis}
   \begin{center}
   \resizebox{.80\linewidth}{!}{
      \begin{tabular}{l|cc |cc | ccccc}
         \toprule
         \multirow{2}{*}{Methods} & \multicolumn{2}{c|}{Train} & \multicolumn{2}{c|}{Test} 
         &\multicolumn{5}{c}{Diagnosis, \%} \\
         \cmidrule(lr){2-10}
         & H. & G. & H. & G. & AUC & Accuracy & Sensitivity & Specificity & F1-score \\
         \midrule
         ABMIL \cite{ilse2018attention} & $\checkmark$ & & $\checkmark$ & & 78.48±1.67 & 55.81±3.26 & 40.09±1.34 & 82.93±0.87 & 35.45±5.48 \\
         TransMIL \cite{shao2021transmil} & $\checkmark$ & & $\checkmark$ & & 79.55±0.39 & 57.27±0.81 & 45.96±4.12 & 85.56±1.23 & 42.37±4.76 \\
         CLAM-SB \cite{lu2021data} & $\checkmark$ & & $\checkmark$ & & 79.39±1.02 & 57.00±2.49 &  45.92±1.85 & 85.74±0.80 & 41.85±4.39 \\
         CLAM-MB \cite{lu2021data} & $\checkmark$ & & $\checkmark$ & & 79.30±1.00 & 56.67±2.05 &  45.38±0.98 & 85.65±0.65 & 41.33±3.62 \\
         DTFD-MIL \cite{zhang2022dtfd} & $\checkmark$ & & $\checkmark$ & & 80.04±0.96 & 59.54±2.87 &  46.46±1.39 & 86.15±0.64 & 42.75±3.20 \\
         DAS-MIL \cite{bontempo2023graph} & $\checkmark$ & & $\checkmark$ & & 78.92±0.96 & 61.46±1.07 &  41.07±1.06 & 85.55±1.03 & 33.75±0.74 \\
         WiKG \cite{li2024dynamic} & $\checkmark$ & & $\checkmark$ & & \underline{83.26±1.68} & \underline{63.25±3.51} &  \underline{48.57±1.83} & \underline{86.69±0.86} & \underline{45.96±1.68} \\
         Ours\ (Stu) & $\checkmark$ & & $\checkmark$ & & \textbf{84.30±2.45} & \textbf{63.90±3.77} & \textbf{54.04±4.14} & \textbf{88.22±1.03} & \textbf{53.25±4.10} \\
         \midrule
         LM & $\checkmark$ & $\checkmark$ & $\checkmark$ & & 79.19±0.99 & 55.83±1.61 & 46.25±1.98 & 85.67±0.70 & 41.58±4.54 \\
         AE \cite{dumpala2019audio} & $\checkmark$ & $\checkmark$ & $\checkmark$ & & 83.95±2.15 & 61.83±3.55 & \underline{51.94±3.15} & 87.84±0.90 & \underline{50.89±2.98} \\
         G-HANet \cite{wang2025histo} & $\checkmark$ & $\checkmark$ & $\checkmark$ & & 83.53±2.25 & 61.75±3.09 & 50.93±2.65 & 87.55±0.80 & 50.17±2.80 \\
         LD-CVAE \cite{zhou2025robust} & $\checkmark$ & $\checkmark$ & $\checkmark$ & & \underline{84.27±0.46} & \underline{64.28±1.91} & 51.02±1.35 & \underline{87.84±0.43} & 48.37±0.07 \\
         Ours\ (Dst) & $\checkmark$ & $\checkmark$ & $\checkmark$ & & \textbf{86.68±1.86} & \textbf{67.39±4.39} & \textbf{55.21±4.06} & \textbf{88.77±1.29} & \textbf{54.85±4.34} \\
         \midrule
         SNN \cite{klambauer2017self} & & $\checkmark$ & & $\checkmark$ & 88.24±0.87 & 73.54±1.35 & 62.59±1.95 & 91.45±0.45 & 61.11±2.97 \\
         Concat & $\checkmark$ & $\checkmark$ & $\checkmark$ & $\checkmark$ & 89.65±1.70 & 73.05±5.18 & 61.90±6.17 & 91.48±1.44 & 61.26±6.47 \\
         Add & $\checkmark$ & $\checkmark$ & $\checkmark$ & $\checkmark$ & 92.28±2.08 & 80.09±3.19 & 71.89±4.30 & 93.47±1.26 & 71.18±4.21 \\
         Bilinear & $\checkmark$ & $\checkmark$ & $\checkmark$ & $\checkmark$ & 94.99±1.07 & 84.64±0.60 & \underline{77.22±1.23} & 95.06±0.27 & 76.38±1.45 \\
         MCAT \cite{chen2021multimodal} & $\checkmark$ & $\checkmark$ & $\checkmark$ & $\checkmark$ & 94.90±1.74 & 82.37±4.25 & 71.26±3.58 & 94.52±1.42 & 70.35±3.21 \\
         CMTA \cite{zhou2023cross} & $\checkmark$ & $\checkmark$ & $\checkmark$ & $\checkmark$ & 89.25±4.07 & 73.44±7.87 & 58.39±10.68 & 91.63±2.36 & 52.49±14.77 \\
         MoME \cite{xiong2024mome} & $\checkmark$ & $\checkmark$ & $\checkmark$ & $\checkmark$ & 93.04±1.02 & 80.83±3.29 & 67.32±1.46 & 93.93±0.94 & 63.18±2.06 \\
         MOTCat \cite{xu2023multimodal} & $\checkmark$ & $\checkmark$ & $\checkmark$ & $\checkmark$ & 93.20±0.39 & 80.88±0.99 & 66.56±2.27 & 93.72±0.31 & 61.43±2.17 \\
         SurvPath \cite{jaume2024modeling} & $\checkmark$ & $\checkmark$ & $\checkmark$ & $\checkmark$ & 94.42±1.33 & 77.96±2.63 & 64.13±8.24 & 91.16±1.58 & 63.07±8.77 \\
         MRePath \cite{qu2025multimodal} & $\checkmark$ & $\checkmark$ & $\checkmark$ & $\checkmark$ & 95.11±0.56 & 82.09±0.49 & 73.11±1.14 & 94.47±0.21 & 72.36±1.88 \\
         MMP \cite{song2024multimodal} & $\checkmark$ & $\checkmark$ & $\checkmark$ & $\checkmark$ & \underline{96.17±0.17} & 85.01±0.68 & 75.82±1.55 & 95.30±0.33 & 74.67±1.73 \\
         SML \cite{zhang2024knowledge} & $\checkmark$ & $\checkmark$ & $\checkmark$ & $\checkmark$ & 96.02±0.38 & \underline{85.52±1.99} & \textbf{77.71±1.45} & \underline{95.58±0.68} & \textbf{77.29±1.31} \\
         Ours\ (Tch) & $\checkmark$ & $\checkmark$ & $\checkmark$ & $\checkmark$ & \textbf{96.31±0.79} & \textbf{86.17±0.90} & 76.66±2.82 & \textbf{95.59±0.22} & \underline{76.40±2.97} \\
         \bottomrule
      \end{tabular}
   }
   \end{center}
\end{table*}
\begin{table*}[!t]
 \scriptsize
    \caption{Comparison with SOTA methods on Grading task (3-fold validation).} \label{table_grading}
   \begin{center}
   \resizebox{.80\linewidth}{!}{
      \begin{tabular}{l |cc |cc |ccccc}
         \toprule
         \multirow{2}{*}{Methods} & \multicolumn{2}{c|}{Train} & \multicolumn{2}{c|}{Test} 
         & \multicolumn{5}{c}{Grading, \%} \\
         \cmidrule(lr){2-10}
         & H. & G. & H. & G. & AUC & Accuracy & Sensitivity & Specificity & F1-score \\
         \midrule
         ABMIL \cite{ilse2018attention} & $\checkmark$ & & $\checkmark$ & & 84.40±0.40 & 64.89±1.32 & 62.81±1.64 & 83.12±0.43 & 62.26±1.14 \\
         TransMIL \cite{shao2021transmil}  & $\checkmark$ & & $\checkmark$ & & 85.73±0.68 & 65.85±2.88 & 61.95±1.27 & 83.36±0.84 & 54.51±5.31 \\
         CLAM-SB \cite{lu2021data} & $\checkmark$ & & $\checkmark$ & & 84.35±0.51 & 66.92±0.81 &  64.76±1.23 & 84.15±0.37 & 63.94±0.35 \\
         CLAM-MB \cite{lu2021data} & $\checkmark$ & & $\checkmark$ & & 84.17±0.50 & 67.02±0.37 &  64.86±0.77 & 84.18±0.15 & 63.94±0.73 \\
         DTFD-MIL \cite{zhang2022dtfd} & $\checkmark$ & & $\checkmark$ & & 84.92±0.44 & 68.20±1.06 &  65.90±1.33 & 84.70±0.46 & 65.25±1.02 \\
         WiKG \cite{li2024dynamic}  & $\checkmark$ & & $\checkmark$ & & \underline{86.94±0.05} & \underline{71.34±0.55} & \underline{68.01±1.31} & \underline{86.01±0.35} & \underline{67.05±1.29} \\
         Ours\ (Stu) & $\checkmark$ & & $\checkmark$ & & \textbf{88.18±0.96} & \textbf{73.45±2.35} & \textbf{70.95±2.63} & \textbf{87.23±1.17} & \textbf{70.38±2.49} \\
         \midrule
         LM & $\checkmark$ & $\checkmark$ & $\checkmark$ & & 84.37±0.55 & 67.46±1.37 & 65.15±1.83 & 84.35±0.60 & 64.33±0.84 \\
         AE \cite{dumpala2019audio} & $\checkmark$ & $\checkmark$ & $\checkmark$ & & 86.87±0.73 & 71.22±1.12 & 68.42±1.22 & 86.09±0.58 & 67.83±1.18 \\ 
         G-HANet \cite{wang2025histo} & $\checkmark$ & $\checkmark$ & $\checkmark$ & & 87.15±0.86 & 71.35±2.39 & 67.82±2.66 & 85.97±1.36 & 67.39±2.25 \\
         LD-CVAE \cite{zhou2025robust} & $\checkmark$ & $\checkmark$ & $\checkmark$ & & \underline{88.42±0.68} & \underline{72.57±1.85} & \underline{69.14±1.93} & \underline{86.59±1.16} & \underline{68.71±2.48} \\
         Ours\ (Dst) & $\checkmark$ & $\checkmark$ & $\checkmark$ & & \textbf{88.56±0.55} & \textbf{74.38±1.46} & \textbf{71.50±1.65} & \textbf{87.57±0.76} & \textbf{70.93±1.41} \\
         \midrule
         SNN \cite{klambauer2017self} & & $\checkmark$ & & $\checkmark$ & 86.79±1.30 & 69.41±1.45 & 66.17±0.97 & 85.35±0.82 & 65.66±1.53 \\
         Concat & $\checkmark$ & $\checkmark$ & $\checkmark$ & $\checkmark$ & 86.94±2.49 & 71.00±3.36 & 67.01±2.65 & 85.89±1.72 & 65.43±3.45 \\
         Add & $\checkmark$ & $\checkmark$ & $\checkmark$ & $\checkmark$ & 84.60±2.22 & 66.08±4.26 & 61.88±3.95 & 83.42±2.31 & 61.28±4.49 \\
         Bilinear & $\checkmark$ & $\checkmark$ & $\checkmark$ & $\checkmark$ & 86.77±0.73 & 70.44±2.51 & 65.75±1.76 & 85.18±1.26 & 64.16±2.10 \\
         MCAT \cite{chen2021multimodal} & $\checkmark$ & $\checkmark$ & $\checkmark$ & $\checkmark$ & 86.74±0.61 & 65.25±4.77 & 62.00±2.85 & 83.33±1.92 & 57.19±5.30 \\
         CMTA \cite{zhou2023cross} & $\checkmark$ & $\checkmark$ & $\checkmark$ & $\checkmark$ & 88.02±1.76 & 71.73±1.23 & 67.90±0.57 & 86.51±0.64 & 63.90±1.00 \\
         MoME \cite{xiong2024mome} & $\checkmark$ & $\checkmark$ & $\checkmark$ & $\checkmark$ & 88.35±1.66 & 72.72±2.77 & 69.06±1.66 & 86.76±1.35 & 66.40±1.52 \\
         MOTCat \cite{xu2023multimodal} & $\checkmark$ & $\checkmark$ & $\checkmark$ & $\checkmark$ & 88.88±0.57 & 72.85±1.92 & 68.59±1.70 & 86.74±0.94 & 64.16±2.58 \\
         MRePath \cite{qu2025multimodal} & $\checkmark$ & $\checkmark$ & $\checkmark$ & $\checkmark$ & 88.70±2.22 & 72.14±3.80 & 68.37±3.27 & 86.43±1.94 & 66.13±3.87 \\
         MMP \cite{song2024multimodal} & $\checkmark$ & $\checkmark$ & $\checkmark$ & $\checkmark$ & \underline{89.12±1.97} & 71.18±2.54 & 67.64±2.79 & 85.97±1.35 & 64.74±1.75 \\
         SML \cite{zhang2024knowledge} & $\checkmark$ & $\checkmark$ & $\checkmark$ & $\checkmark$ & 88.37±2.23 & \underline{73.71±3.58} & \underline{70.54±2.80} & \underline{87.40±1.73} & \underline{69.12±3.51} \\
         Ours\ (Tch) & $\checkmark$ & $\checkmark$ & $\checkmark$ & $\checkmark$ & \textbf{89.15±1.64} & \textbf{74.93±3.73} & \textbf{71.21±3.10} & \textbf{87.79±1.79} & \textbf{69.99±3.29} \\
         \bottomrule
      \end{tabular}
   }
   \end{center}
\end{table*}

\subsubsection{Comparisons}
For downstream tasks, we compared our model against twenty-four SOTA methods. i) WSI-based methods: ABMIL \cite{ilse2018attention}, TransMIL \cite{shao2021transmil}, CLAM-SB \cite{lu2021data}, CLAM-MB \cite{lu2021data}, DTFD-MIL \cite{zhang2022dtfd}, DAS-MIL \cite{bontempo2023graph} (graph-based multi-scale approach), WiKG \cite{li2024dynamic} (graph-based approach); ii) Transcriptomic-based method: SNN \cite{klambauer2017self}); iii) Multi-modal methods: Concat (ABMIL with SNN), Add (ABMIL with SNN), Bilinear (ABMIL with SNN), MCAT \cite{chen2021multimodal}, CMTA \cite{zhou2023cross}, MoME \cite{xiong2024mome}, MOTCat \cite{xu2023multimodal}, SurvPath\cite{jaume2024modeling}, PIBD\cite{zhang2024prototypical}, MRePath\cite{qu2025multimodal}, MMP \cite{song2024multimodal} (multi-modal prototyping approach), and SML \cite{zhang2024knowledge} (our conference version);  iv) Multi-modal learning methods handling missing modalities: LM (Linear Mapping), AE (Autoencoder-based Imputation) \cite{dumpala2019audio}, G-HANet \cite{wang2025histo} (Genome-informed Hyper-attention Network), and LD-CVAE \cite{zhou2025robust} (Conditional Latent Differentiation Variational
Autoencoder-based Imputation). 
\begin{table}[!t]
 \scriptsize
    \caption{Comparison with SOTA methods on Survival prediction.} \label{table_survival}
   \begin{center}
   \resizebox{.99\linewidth}{!}{
      \begin{tabular}{l|cc |cc |cc}
         \toprule
         \multirow{2}{*}{Methods} & \multicolumn{2}{c|}{Train} & \multicolumn{2}{c|}{Test}
         & Internal & External \\
         \cmidrule(lr){2-7}
         & H. & G. & H. & G. & C-Index, \% & Zero-Shot, \%\\
         \midrule
         ABMIL \cite{ilse2018attention} & $\checkmark$ & & $\checkmark$ & & 67.06±4.01 & 57.36\\
         TransMIL \cite{shao2021transmil} & $\checkmark$ & & $\checkmark$ & & 73.27±4.84 & 59.26\\
         CLAM-SB \cite{lu2021data} & $\checkmark$ & & $\checkmark$ & & 71.85±5.02 & 59.19 \\
         CLAM-MB \cite{lu2021data} & $\checkmark$ & & $\checkmark$ & & 71.81±4.96 & 59.43 \\
         DTFD-MIL \cite{zhang2022dtfd} & $\checkmark$ & & $\checkmark$ & & 71.89±4.95 & \underline{59.48} \\
         DAS-MIL \cite{bontempo2023graph} & $\checkmark$ & & $\checkmark$ & & 71.84±4.94 & 59.21 \\
         WiKG \cite{li2024dynamic} & $\checkmark$ & & $\checkmark$ & & \underline{73.95±4.81} & 58.02 \\
         Ours\ (Stu) & $\checkmark$ & & $\checkmark$ & & \textbf{73.98±4.29} & \textbf{60.15} \\
         \midrule
         LM & $\checkmark$ & $\checkmark$ & $\checkmark$ & & 71.81±5.15 & 59.01 \\
         AE \cite{dumpala2019audio} & $\checkmark$ & $\checkmark$ & $\checkmark$ & & 73.76±4.94 & \underline{59.31} \\
         G-HANet \cite{wang2025histo} & $\checkmark$ & $\checkmark$ & $\checkmark$ & & 73.84±4.38 & 57.96 \\
         LD-CVAE \cite{zhou2025robust} & $\checkmark$ & $\checkmark$ & $\checkmark$ & & \underline{74.01±3.78} & 56.67\\
         Ours\ (Dst) & $\checkmark$ & $\checkmark$ & $\checkmark$ & & \textbf{74.47±3.93} & \textbf{59.63}\\
         \midrule
         SNN \cite{klambauer2017self} & & $\checkmark$ & & $\checkmark$ & 75.99±5.32 & 54.54\\
         Concat & $\checkmark$ & $\checkmark$ & $\checkmark$ & $\checkmark$ & 76.10±3.85 & 55.49\\
         Add & $\checkmark$ & $\checkmark$ & $\checkmark$ & $\checkmark$ & 73.99±1.67 & 50.00\\
         Bilinear & $\checkmark$ & $\checkmark$ & $\checkmark$ & $\checkmark$ & 76.31±2.82 & 53.73\\
         MCAT \cite{chen2021multimodal} & $\checkmark$ & $\checkmark$ & $\checkmark$ & $\checkmark$ & 75.01±4.62 & 60.43\\
         CMTA \cite{zhou2023cross} & $\checkmark$ & $\checkmark$ & $\checkmark$ & $\checkmark$ & 75.34±2.37 & 52.64\\
         MoME \cite{xiong2024mome} & $\checkmark$ & $\checkmark$ & $\checkmark$ & $\checkmark$ & 75.32±3.82 & 59.63 \\
         MOTCat \cite{xu2023multimodal} & $\checkmark$ & $\checkmark$ & $\checkmark$ & $\checkmark$ & 76.58±5.60 & 60.66 \\
         SurvPath \cite{jaume2024modeling} & $\checkmark$ & $\checkmark$ & $\checkmark$ & $\checkmark$ & 76.70±3.41 & 55.10 \\
         PIBD \cite{zhang2024prototypical} & $\checkmark$ & $\checkmark$ & $\checkmark$ & $\checkmark$ & 76.38±4.75 & \underline{63.52} \\
         MRePath \cite{qu2025multimodal} & $\checkmark$ & $\checkmark$ & $\checkmark$ & $\checkmark$ & 76.42±2.22 & 53.76 \\
         MMP \cite{song2024multimodal} & $\checkmark$ & $\checkmark$ & $\checkmark$ & $\checkmark$ & \underline{77.10±1.57} & 57.53 \\
         SML \cite{zhang2024knowledge} & $\checkmark$ & $\checkmark$ & $\checkmark$ & $\checkmark$ & 76.55±2.10 & 55.53\\
         Ours\ (Tch) & $\checkmark$ & $\checkmark$ & $\checkmark$ & $\checkmark$ & \textbf{77.49±2.57} & \textbf{65.18}\\
         \bottomrule
      \end{tabular}
   }
   \end{center}
\end{table}
\subsection{Implementation Details}
\begin{figure}[!t]
\centering
\includegraphics[scale=.30]{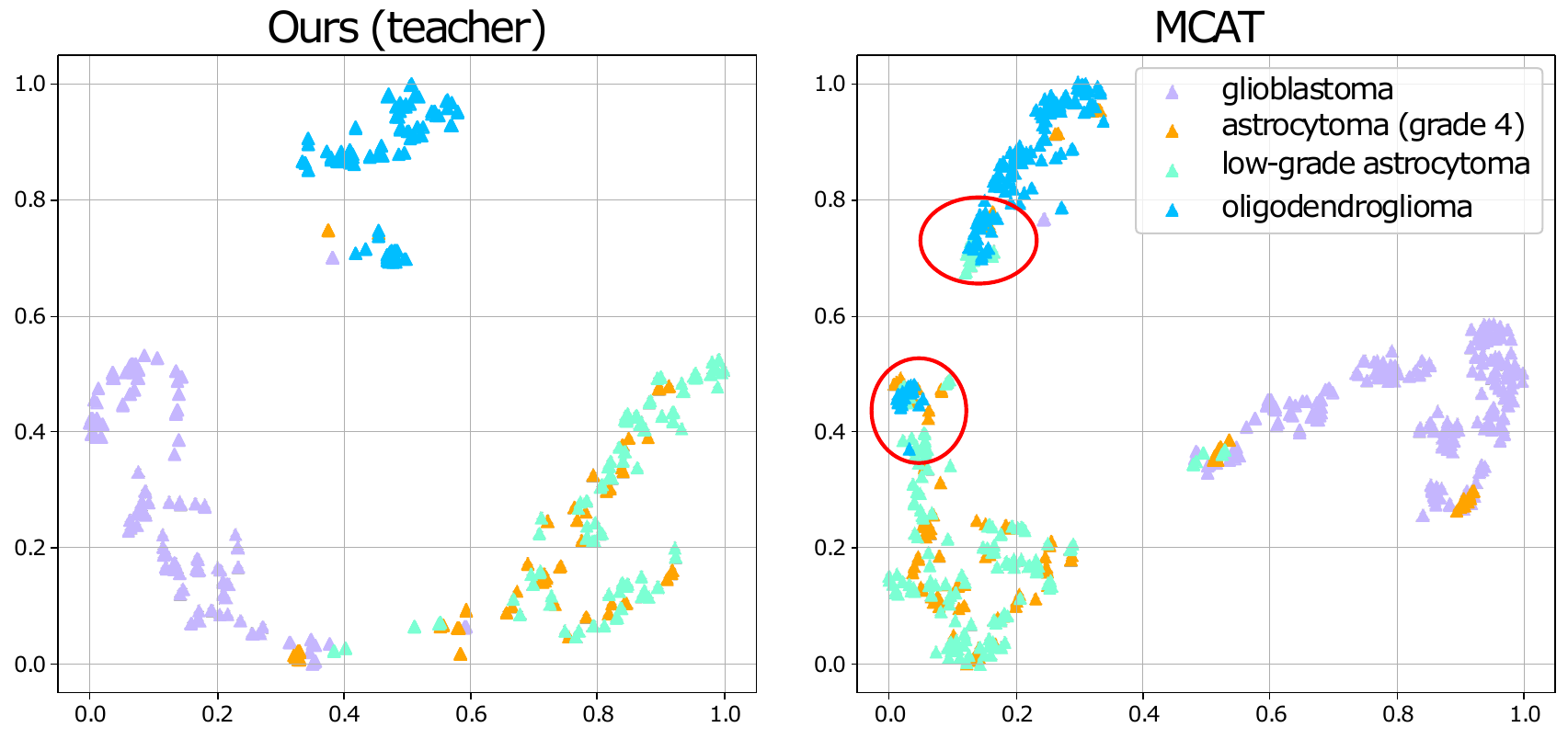} 
\caption{Visualization of feature representation using Ours (teacher) and MCAT in glioma diagnosis on TCGA GBM-LGG datasets. Our teacher exhibits more distinct clustering, particularly in distinguishing low-grade astrocytoma and oligodendroglioma cases, as circled in red.} \label{fig:TSNE}
\end{figure}
Each WSI was downsampled to obtain representations at $10\times$ (1$\mu$m $\textrm{px}^{-1}$, tissue level) and $20\times$ (0.5$\mu$m $\textrm{px}^{-1}$, cell level) magnifications. Each magnification was divided into non-overlapping patches of size 224 $\times$ 224 px. Following \cite{wang2023multi}, we sampled 2,500 patches per WSI using a biologically informed repeat strategy to ensure representative coverage. Color normalization \cite{vahadane2016structure} was used to reduce staining variability. 

Patch features extracted using a ResNet50 pre-trained on ImageNet \cite{deng2009imagenet} were concatenated into slide-level feature matrices for downstream processing.
For transcriptomics, we followed \cite{bhattacharya2018immport} to identify shared signatures in the TCGA \cite{tomczak2015review} and IvyGAP \cite{puchalski2018anatomic} datasets. According to finding 2 (Section \ref{method:findings2}.2), we selected the top 30\% of Highly Variable Genes (HVGs: genes with a high signal-to-noise ratio, enabling a compact and generalizable representation of the transcriptome), which capture biologically informative variation.
We utilized the same preprocessing for all the baselines, including patch sampling and gene selection. For the pathway-based methods, we followed the original papers and utilized all genes needed.
All experiments were implemented using  PyTorch \cite{paszke2019pytorch} on two NVIDIA RTX A5000 GPUs. We employed 3-fold cross-validation across all downstream tasks, training for 10 epochs per fold, and optimized parameters using the AdamW optimizer \cite{kingma2014adam} with tuned hyperparameters. 
\begin{table*}[!t]
 \scriptsize
    \caption{Ablation studies on Diagnosis, Grading, and Survival tasks at Stage I. The best results are highlighted with \textbf{bold}.} \label{table_ablation_stageI}
   \begin{center}
   \resizebox{.99\linewidth}{!}{ 
      \begin{tabular}{lccc | ccccc | ccc | c}
         \toprule
         
         &\multicolumn{3}{c|}{Methods}
         &\multicolumn{5}{c|}{Diagnosis, \%}   
         &\multicolumn{3}{c|}{Grading, \%}
         &Survival, \%
         \\
         
         \cmidrule(lr){2-13}  

          & CGC   & IGC   & GS-Layer
          & AUC   & Accuracy   & Sensitivity   & Specificity   &F1-score   
          & AUC   & Accuracy   &F1-score& C-Index\\
          
         \midrule

         1           &   &   &
         &93.73±1.32 &79.32±3.62 &69.31±3.24 &93.43±1.22 &67.92±5.21 
         &87.71±1.58 &69.42±2.59 &60.48±2.87 &75.82±2.88\\
         
         2           & \checkmark &   &
         &96.03±0.80 &84.64±1.29 &75.07±2.75 &95.14±0.60 &74.75±2.54 
         &88.52±1.93 &71.18±3.32 &63.21±1.60 &76.41±2.19\\
         
         3           &           &\checkmark &
         &95.30±0.52 &82.98±1.07 &73.57±2.78 &94.57±0.40 &72.72±2.83 
         &88.47±1.90 &71.41±3.43 &63.43±1.41 &76.25±3.38\\

         4           &           &    &\checkmark
         &94.79±0.51 &82.97±0.98 &74.77±1.22 &94.68±0.30 &73.66±1.57 
         &88.73±1.55 &72.95±2.84 &65.19±3.24 &75.85±2.07\\
         
         5            &\checkmark &   &\checkmark
         &95.51±0.60 &84.30±1.19 &76.06±0.97 &95.00±0.49        &75.50±0.97
         &89.09±2.14 &74.65±4.65 &70.54±5.01
         &76.14±2.30\\
         
         6      &   & \checkmark  &\checkmark
         &95.55±0.59 &84.24±1.26 &75.82±1.03 &94.98±0.51 &75.29±1.08 
         &89.06±2.05 &74.35±4.42 &70.14±4.70 
         &76.42±2.34\\

         7           &\checkmark & \checkmark &
         &96.07±0.79 &85.00±1.58 &76.23±2.94 &95.28±0.70 &75.78±2.58 
         &89.13±2.22 &74.41±2.95 &\textbf{70.69±3.02} 
         &76.43±3.25\\
         
         8 &\checkmark &\checkmark &\checkmark
         &\textbf{96.31±0.79} &\textbf{86.17±0.90} &\textbf{76.66±2.82} &\textbf{95.59±0.22} &\textbf{76.40±2.97} 
         &\textbf{89.15±1.64} &\textbf{74.93±3.73} 
         &69.99±3.29 
         &\textbf{77.49±2.57}\\
         
         \bottomrule
      \end{tabular}
    }
   \end{center}
\end{table*}
\begin{table*}[ht]
\centering
\caption{Ablation studies of Stage II across three downstream tasks (Diagnosis, Grading, Survival)}
\label{table_ablation_stageII}
\resizebox{.999\linewidth}{!}{ 
\begin{tabular}{ll|ccccc|ccccc|c}
\toprule
\multicolumn{2}{c|}{\multirow{2}{*}{Methods}} 
& \multicolumn{5}{c|}{Diagnosis} 
& \multicolumn{5}{c|}{Grading}  
& \multicolumn{1}{c}{Survival} \\
\cmidrule(lr){3-13}
\multicolumn{2}{c|}{} 
& AUC & ACC & Sens & Spec & F1
& AUC & ACC & Sens & Spec & F1
& C-Index \\
\cmidrule(lr){1-13}

\multirow{6}{*}{ITA} 
&$w/o$ H-to-H Deformation                & 82.28±1.70 & 58.97±3.52 & 48.66±1.85 & 86.40±0.58 & 46.27±1.80 & 86.43±1.14 & 69.03±2.56 & 66.31±2.23 & 85.01±0.91 & 66.05±1.97 & 72.17 $\pm$ 4.56 \\
&replace H-to-H with Self-Att            & 82.82±1.63 & 60.17±5.09 & 50.78±3.07 & 87.03±0.75 & 48.92±4.80 & 86.28±1.63 & 69.73±3.64 & 67.27±3.41 & 85.36±1.45 & 6654±2.72 & 72.72 $\pm$ 4.77 \\
\cmidrule(lr){2-13}
&$w/o$ Clustering and Merging            & 82.63±2.04 & 60.37±3.04 & 49.73±1.25 & 87.00±0.63 & 48.99±1.21 & 86.97±0.31 & 67.38±1.27 & 64.55±1.56 & 84.30±0.47 & 62.40±3.11 & 73.67 $\pm$ 4.17 \\
&replace 2 with 8 clusters               & 84.27±1.59 & 62.16±3.68 & 47.88±5.88 & 85.98±2.45 & 45.01±7.84 & 87.34±0.22 & 69.25±1.13 & 66.00±1.43 & 85.09±0.65 & 64.30±1.33 & 73.08 $\pm$ 4.36 \\
&replace 2 with 32 clusters              & 84.02±2.62 & 63.81±4.69 & 51.57±3.22 & 87.59±1.09 & 50.18±3.34 & 87.19±0.88 & 68.96±2.33 & 65.53±2.01 & 84.88±1.04 & 63.92±1.63 & 73.58 $\pm$ 4.74 \\
\cmidrule(lr){2-13}
&Ours (Stu)                    & \textbf{84.30±2.45} & \textbf{63.90±3.77} & \textbf{54.04±4.14} & \textbf{88.22±1.03} & \textbf{53.25±4.10} & \textbf{88.18±0.96} & \textbf{73.45±2.35} & \textbf{70.95±2.63} & \textbf{87.23±1.17} & \textbf{70.38±2.49} & \textbf{73.98 $\pm$ 4.29} \\ 
\cmidrule(lr){1-13}

\multirow{3}{*}{SKD} 
&$w/o$ MSE loss                         & 84.78±2.58 & 65.26±4.64 & 52.86±3.50 & 87.88±1.15 & 51.08±4.05 & 88.39±0.47 & 72.62±0.98 & 69.53±1.09 & 86.71±0.36 & 68.63±0.88 & 74.31 $\pm$ 4.84 \\
&$w/o$ KL loss                          & 84.40±2.36 & 61.62±3.90 & 52.23±4.25 & 87.53±1.50 & 49.68±4.76 & 88.09±0.90 & 71.65±1.73 & 68.53±1.75 & 86.23±0.78 & 67.34±1.96 & 74.38 $\pm$ 4.45 \\
&Ours (Dst)                             & \textbf{86.68±1.86} & \textbf{67.39±4.39} & \textbf{55.21±4.06} & \textbf{88.77±1.29} & \textbf{54.85±4.34} & \textbf{88.56±0.55} & \textbf{74.38±1.46} & \textbf{71.50±1.65} & \textbf{87.57±0.76} & \textbf{70.93±1.41} & \textbf{74.47 $\pm$ 3.93} \\
\bottomrule
\end{tabular}
}
\end{table*}

\begin{table}[ht]
\centering
\caption{Ablation studies on different input magnifications}
\label{table_magnification}
\resizebox{.99\linewidth}{!}{ 
\begin{tabular}{l|c|ccccc}
\toprule
\multicolumn{1}{c|}{Task} & \multicolumn{1}{c|}{Magnification} & AUC & ACC & Sens & Spec & F1 \\
\cmidrule(lr){1-7} 
\multirow{3}{*}{Diag} 
& 10$\times$   & 95.87 $\pm$ 0.38 & 84.53 $\pm$ 0.91 & 74.97 $\pm$ 1.06 & 95.11 $\pm$ 0.24 & 73.36 $\pm$ 2.09 \\
& 20$\times$   & 95.83 $\pm$ 0.38 & 84.33 $\pm$ 0.94 & 75.08 $\pm$ 2.04 & 95.10 $\pm$ 0.49 & 73.72 $\pm$ 2.36 \\
& Ours   & \textbf{96.31 $\pm$ 0.79} & \textbf{86.17 $\pm$ 0.90} & \textbf{76.66 $\pm$ 2.82} & \textbf{95.59 $\pm$ 0.22} & \textbf{76.40 $\pm$ 2.97} \\
\cmidrule(lr){1-7} 
\multirow{3}{*}{Grad} 
& 10$\times$   & 88.66 $\pm$ 2.20 & 73.66 $\pm$ 4.21 & 69.85 $\pm$ 3.82 & 87.18 $\pm$ 2.10 & 68.18 $\pm$ 4.73 \\
& 20$\times$   & 88.61 $\pm$ 2.27 & 73.70 $\pm$ 4.07 & 70.01 $\pm$ 3.52 & 87.25 $\pm$ 1.99 & 68.58 $\pm$ 4.20 \\
& Ours   & \textbf{89.15 $\pm$ 1.64} & \textbf{74.93 $\pm$ 3.73} & \textbf{71.21 $\pm$ 3.10} & \textbf{87.79 $\pm$ 1.79} & \textbf{69.99 $\pm$ 3.29} \\
\bottomrule
\end{tabular}
}
\end{table}

\begin{table}[ht]
\centering
\caption{Ablation studies on gene disentanglement}
\label{table_singleGeneGroup}
\resizebox{.99\linewidth}{!}{ 
\begin{tabular}{l|c|ccccc}
\toprule
\multicolumn{1}{c|}{Task} & \multicolumn{1}{c|}{Methods} & AUC & ACC & Sens & Spec & F1 \\
\cmidrule(lr){1-7} 
\multirow{4}{*}{Diag} 
&$w/o$ Tumor   & 95.69 $\pm$ 0.39 & 84.66 $\pm$ 0.74 & 75.02 $\pm$ 2.21 & 95.10 $\pm$ 0.21 & 73.75 $\pm$ 2.77 \\
&$w/o$ TME   & 88.47 $\pm$ 0.90 & 71.85 $\pm$ 2.90 & 57.25 $\pm$ 3.57 & 90.04 $\pm$ 0.75 & 56.57 $\pm$ 3.49 \\
&$w/o$ disen   & 95.90 $\pm$ 0.18 & 85.71 $\pm$ 0.15 & 76.72 $\pm$ 0.34 & 95.52 $\pm$ 0.26 & 75.75 $\pm$ 0.33 \\
&Ours   & \textbf{96.31 $\pm$ 0.79} & \textbf{86.17 $\pm$ 0.90} & \textbf{76.66 $\pm$ 2.82} & \textbf{95.59 $\pm$ 0.22} & \textbf{76.40 $\pm$ 2.97} \\
\cmidrule(lr){1-7} 
\multirow{4}{*}{Grad} 
&$w/o$ Tumor   & 88.51 $\pm$ 2.14 & 73.72 $\pm$ 3.13 & 70.71 $\pm$ 2.68 & 87.39 $\pm$ 1.60 & 69.16 $\pm$ 3.98 \\
&$w/o$ TME   & 84.13 $\pm$ 0.62 & 66.56 $\pm$ 2.28 & 63.58 $\pm$ 2.37 & 83.83 $\pm$ 0.99 & 61.93 $\pm$ 3.93 \\
&$w/o$ disen   & 89.05 $\pm$ 2.02 & 73.35 $\pm$ 3.95 & 70.28 $\pm$ 3.86 & 87.18 $\pm$ 1.99 & 68.54 $\pm$ 5.14 \\
&Ours   & \textbf{89.15 $\pm$ 1.64} & \textbf{74.93 $\pm$ 3.73} & \textbf{71.21 $\pm$ 3.10} & \textbf{87.79 $\pm$ 1.79} & \textbf{69.99 $\pm$ 3.29} \\
\bottomrule
\end{tabular}
}
\end{table}

\begin{table}[ht]
\centering
\caption{Ablation studies on survival prediction with teacher model}
\label{table_survAblation}
\resizebox{.99\linewidth}{!}{ 
\begin{tabular}{l|cccccc}
\toprule
Metric & 10 $\times$ & 20 $\times$ & $w/o$ disen & $w/o$ Tumor & $w/o$ TME & Ours \\ 
\cmidrule(lr){1-7} 
C-Index & 76.95 $\pm$ 3.30 & 76.75 $\pm$ 3.55 & 76.74 $\pm$ 2.40 & 76.37 $\pm$ 1.91 & 76.82 $\pm$ 3.69 & \textbf{77.49 $\pm$ 2.57} \\
\bottomrule
\end{tabular}
}
\end{table}

\begin{table}[ht]
\centering
\caption{Performance comparison of different input gene selection for glioma diagnosis}
\label{table_hvgs}
\resizebox{.99\linewidth}{!}{ 
\begin{tabular}{l|ccccc}
\toprule
Metric & 10\% HVGs & 30\% HVGs & 30\% Random & 50\% HVGs & 80\% HVGs \\ 
\cmidrule(lr){1-6} 
AUC & 94.33 $\pm$ 0.46 & \textbf{96.31 $\pm$ 0.79} & 95.72 $\pm$ 0.46 & 96.27 $\pm$ 0.96 & 96.11 $\pm$ 0.49 \\
ACC & 80.90 $\pm$ 0.93 & \textbf{86.17 $\pm$ 0.90} & 83.24 $\pm$ 1.63 & 84.74 $\pm$ 2.27 & 85.67 $\pm$ 0.26 \\
Sens & 69.55 $\pm$ 1.04 & \textbf{76.66 $\pm$ 2.82} & 71.92 $\pm$ 2.95 & 75.40 $\pm$ 2.01 & 74.37 $\pm$ 0.66 \\
Spec & 93.39 $\pm$ 0.41 & \textbf{95.59 $\pm$ 0.22} & 94.23 $\pm$ 0.44 & 95.04 $\pm$ 0.87 & 95.12 $\pm$ 0.29 \\
F1 & 69.20 $\pm$ 2.03 & \textbf{76.40 $\pm$ 2.97} & 72.20 $\pm$ 3.16 & 74.43 $\pm$ 1.78 & 73.53 $\pm$ 1.64 \\
\bottomrule
\end{tabular}
}
\end{table}

\begin{table}[ht]
\centering
\caption{Performance comparison of different k in knn for glioma diagnosis with our student model}
\label{table_knn}
\resizebox{.99\linewidth}{!}{ 
\begin{tabular}{l|ccccc}
\toprule
Metric & k = 3 & k = 5 & k = 7 & k = 9 & k = 11 \\ 
\cmidrule(lr){1-6} 
AUC & 83.76 $\pm$ 1.22 & \textbf{84.30 $\pm$ 2.45} & 83.86 $\pm$ 1.38 & 83.70 $\pm$ 1.29 & 83.73 $\pm$ 1.26 \\
ACC & 62.43 $\pm$ 2.86 & \textbf{63.90 $\pm$ 3.77} & 62.38 $\pm$ 3.07 & 62.27 $\pm$ 2.82 & 62.38 $\pm$ 3.25 \\
Sens & 50.70 $\pm$ 1.72 & \textbf{54.04 $\pm$ 4.14} & 50.75 $\pm$ 2.08 & 50.51 $\pm$ 1.56 & 50.69 $\pm$ 1.93 \\
Spec & 87.06 $\pm$ 0.63 & \textbf{88.22 $\pm$ 1.03} & 87.09 $\pm$ 0.62 & 87.00 $\pm$ 0.58 & 87.06 $\pm$ 0.54 \\
F1 & 48.55 $\pm$ 3.20 & \textbf{53.25 $\pm$ 4.10} & 48.72 $\pm$ 3.72 & 48.45 $\pm$ 2.87 & 48.62 $\pm$ 3.74 \\
\bottomrule
\end{tabular}
}
\end{table}

\subsection{Experimental Results}
\subsubsection{Glioma Diagnosis}
Table~\ref{table_diagnosis} presents glioma diagnosis performance under different experimental settings.
In the \textbf{uni-modal} setting, our student model (Ours (Stu)) achieves an AUC of 84.30$\pm$2.45\%, outperforming prior methods. For instance, it is 1.04\% higher than WiKG in AUC. Notably, our student model surpasses WiKG by 7.29\% in F1-score,  further confirming the superiority of our model in WSI-only settings.

In the \textbf{missing-modality} setting, our distillation model (Ours (Dst)), trained with multi-modal knowledge and tested on WSI only, achieves an AUC of 86.68$\pm$1.86\% and outperforms LD-CVAE by 2.41\%. Similarly, it achieves superior accuracy (67.39$\pm$4.39\%) and F1-score (54.85$\pm$4.34\%), with improvements of 3.11\% and 3.96\% over the second best method. This confirms the effectiveness of our distillation in transferring multi-modal knowledge to a single modality.

In the \textbf{multi-modal} setting, our teacher model (Ours (Tch)) achieves the best AUC of 96.31$\pm$0.79\%, demonstrating the effectiveness of our method.  
As illustrated in Fig. \ref{fig:TSNE}, the teacher model demonstrates superior class separability, notably distinguishing oligodendroglioma from low-grade astrocytoma.
These consistent improvements across three settings validate the effectiveness of our framework, highlighting the effectiveness of our distillation approach, narrowing the performance gap between uni-modal and multi-modal models.

\subsubsection{Glioma Grading} 
As shown in Table~\ref{table_grading}, our model consistently achieves the best performance across all metrics under three settings, demonstrating strong robustness and effectiveness. In the \textbf{uni-modal} setting, our student model achieves an AUC of 88.18$\pm$0.96\% and accuracy of 73.45$\pm$2.35\%, outperforming WiKG by 1.24\% and 2.11\%, respectively. The student also attains an F1-score of 70.38$\pm$2.49\%, surpassing all uni-modal and several multi-modal models, highlighting the strength of our ITA-based WSI representation.

In the \textbf{missing-modality} setting, our distillation model achieves an accuracy of 74.38$\pm$1.46\% and F1-score of 70.93$\pm$1.41\%, outperforming the second-best by 1.81\% and 2.22\%, respectively. These results suggest that our distillation is effective in transferring multi-modal knowledge to a single modality for WSI-only inference. 

Finally, in the \textbf{multi-modal} setting, our teacher model achieves best AUC (89.15$\pm$1.64\%), accuracy (74.93$\pm$3.73\%), and F1-score (69.99$\pm$3.29\%), indicating a balanced and robust grading capability. 
Compared to the second-best method (SML), our model shows improvements of 1.22\% in accuracy. 
These results validate the advantages of our model for reliable glioma grading in both multi-modal and WSI-only scenarios.

\subsubsection{Survival Prediction} 
Following prior studies~\cite{chen2020pathomic, chen2021multimodal}, we discretized the overall survival into four time intervals using the quartiles of survival time and evaluated model performance using the discretized-survival C-index. As shown in Table~\ref{table_survival}, our multi-modal teacher achieves the best C-index of 77.49$\pm$2.57\%, outperforming all competing methods. This confirms the effectiveness of integrating transcriptomic and histological cues for survival modeling.
\begin{figure}[!t]
\centering
\includegraphics[scale=.48]{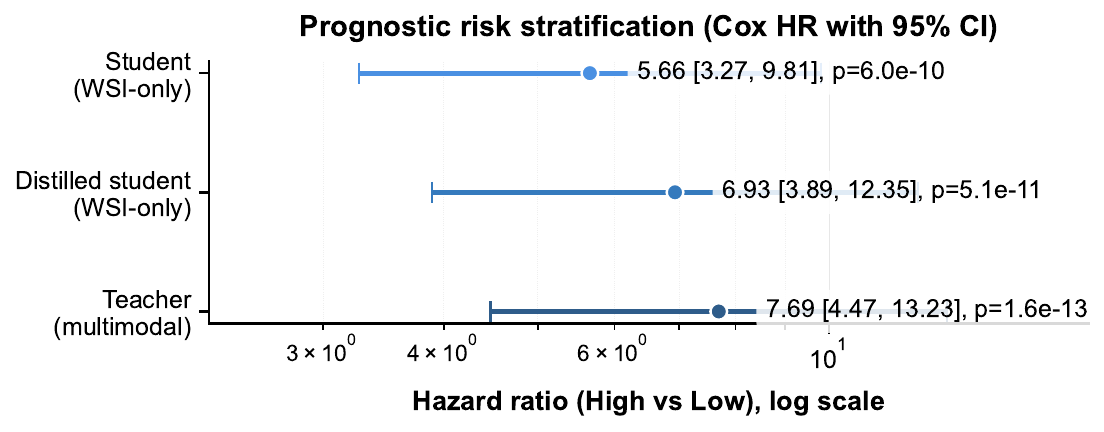} 
\caption{Survival risk stratification measured by Cox hazard ratios. Cox hazard ratios (high vs. low risk) with 95\% confidence intervals summarize the strength of prognostic stratification. Risk groups are defined by a median split of the predicted risk score.} \label{fig:hr}
\end{figure}

In the \textbf{uni-modal} setting, our student model achieves the best C-index of 73.98$\pm$4.29\%, demonstrating that our ITA-enhanced WSI representation can effectively predict survival even without transcriptomic input. In the \textbf{missing-modality} setting, our distilled model yields the best result among all WSI-only inference models, achieving a C-index of 74.47$\pm$3.93\%, surpassing LD-CVAE by 0.46\%, and a \textbf{multi-modal} method (Add, 73.99$\pm$1.67\%). This again validates the effectiveness of our distillation strategy. These results collectively confirm the strength of our framework, including student, diatillation, and teacher models, which provide robust survival prediction under both ideal and real-world settings.

Beyond the C-index, we further extended the survival evaluation by conducting Cox proportional hazards analysis with risk stratification based on predicted risk scores. Patients were dichotomized into high- and low-risk groups using a median split, and hazard ratios with 95\% confidence intervals were estimated to quantify the strength of prognostic separation.  
As shown in Fig.~\ref{fig:hr}, the multi-modal teacher yields the largest hazard ratio, indicating stronger prognostic separation, while the distilled WSI-only student preserves substantial risk stratification without requiring transcriptomic data during inference.

\subsection{Analysis of Our Framework}
\subsubsection{Zero-shot Transfer Evaluation}
\label{subsub:external}
To assess generalizability, we performed a zero-shot transfer experiment by directly applying the pre-trained models to an external cohort (CPTAC) without any fine-tuning. As shown in Table~\ref{table_survival}, our multi-modal teacher model achieves the highest C-index of 65.18\%, outperforming all competing methods.

Our WSI-only student model achieves a C-index of 60.15\%, surpassing all \textbf{uni-modal} baselines, and some \textbf{multi-modal} approaches such as Add (50.00\%), Bilinear (53.73\%), and CMTA (52.64\%). Notably, our distilled model, even without fine-tuning, achieves a C-index of 59.63\%, outperforming \textbf{missing-modality} methods (e.g., G-HANet and LD-CVAE).
These demonstrate that our framework exhibits strong out-of-distribution generalization. In particular, the teacher’s high performance and the student’s competitive zero-shot accuracy confirm that multi-modal subspace learning and distillation preserve essential predictive signals. This highlights the clinical potential of our framework for real-world deployment where transcriptomics may be unavailable and external variability is high.

\subsubsection{Ablation Study}
To quantitatively assess the contribution of our proposed components, we conducted ablation studies on \textbf{i)} model design at two stages, \textbf{ii)} magnification, and \textbf{iii)} gene disentanglement across all downstream tasks.

\textbf{i)} Table~\ref{table_ablation_stageI} shows the ablation results of teacher model at \textbf{Stage I}, compared to the baseline (line 1), adding the CGC strategy (line 2), IGC strategy (line 3), and GS-Layer (line 4) improves diagnosis accuracy by 5.32\%, 3.66\%, and 3.65\%, respectively; and F1-score by 6.83\%, 4.8\%, and 5.74\%, respectively. 
These results confirm that both modules independently enhance the model’s discriminative ability by enforcing biologically informed structure and multi-scale consistency.
Notably, the full model integrating CGC, IGC, and GS-Layer (line 8) achieves the best performance across all tasks. In the diagnosis task, it attains the highest AUC (96.31\%), accuracy (86.17\%), and F1-score (76.40\%). In grading, it yields the highest scores in AUC (89.15\%) and accuracy (74.93\%). For survival prediction, it achieves the C-index of 77.49\%, surpassing the baseline by 1.67\%. These consistent gains across different tasks and metrics validate the effectiveness of each component and confirm their synergistic effect when combined in our full multi-modal teacher framework.
\textbf{Stage II} is designed to distill the proposed ITA-based student model using the SKD strategy. As shown in Table~\ref{table_ablation_stageII}, disabling the H-to-H Deformation module or replacing it with a self-attention layer in the ITA module results in performance degradation across all tasks compared with the full student model (Ours (Stu)). 
For the SKD strategy, removing either the MSE loss or the KL divergence loss leads to reduced performance, thereby demonstrating the effectiveness of the proposed SKD strategy.

\textbf{ii)} Table~\ref{table_magnification} and Table~\ref{table_survAblation} present the ablation results under different magnification settings. Using either $10\times$-only or $20\times$-only inputs leads to reduced performance compared with multi-magnification integration.
Similarly, \textbf{iii)} using Tumor-only or TME-only gene sets results in performance degradation, as shown in Table~\ref{table_singleGeneGroup} and Table~\ref{table_survAblation}.
Consistent with Preliminary Finding~2 (Section~\ref{method:findings2}.2), parallel input of tumor and TME genes (Ours) improves performance compared with treating all genes as a single group without disentanglement (w/o disen).

\subsubsection{Hyper-parameters Sensitivity Analysis}
To evaluate the sensitivity to hyper-parameters, we conduct experiments by varying \textbf{i)} input genes, \textbf{ii)} the number of clusters, and \textbf{iii)} the parameter k in DPC-KNN \cite{du2016study}. 
\textbf{i)} Retaining sufficient biological signal while reducing feature dimensions remains essential for transcriptome modeling. 
The number of HVGs must balance biological informativeness and computational efficiency.
To evaluate the impact of the number of HVGs, we varied the HVG selection threshold across 10\%, 30\%, 50\%, and 80\%.  
As shown in Table~\ref{table_hvgs}, the best performance is observed with 30\% HVGs across all metrics, particularly in AUC (96.31\%) and accuracy (86.17\%). This threshold also outperforms randomly selected 30\% genes, indicating that this HVG selection captures biologically relevant signals while mitigating noise that could hinder classification. 
Despite fluctuations in model performances due to variations in HVGs numbers, our approach consistently outperforms most SOTA models, demonstrating its robustness.
\textbf{ii)} Following previous work \cite{song2024morphological}, we vary the number of clusters from the default setting of two to alternative settings in the Clustering and Merging of ITA module. The results shown in Table \ref{table_ablation_stageII} indicate that our student model is not highly sensitive to the choice of cluster number, with different settings yielding comparable performance across all tasks.
\textbf{iii)} As shown in Table \ref{table_knn}, the diagnosis results under different values of K in DPC-KNN for the student model are generally stable, and the best performance is achieved at $k = 5$.

\subsubsection{Gene-level Interpretability}
\begin{figure}[!t]
\centering
\includegraphics[scale=.19]{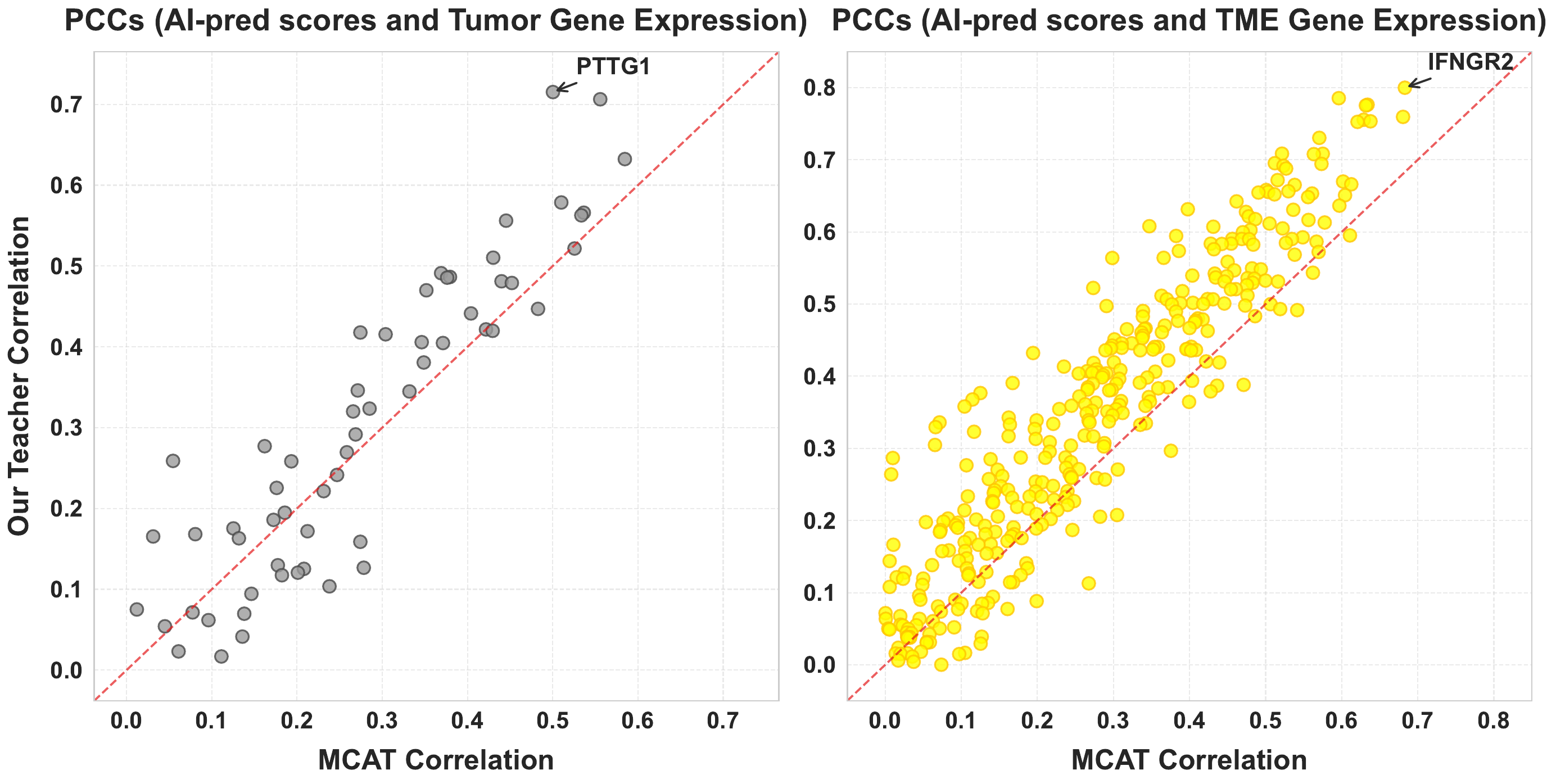} 
\caption{Pearson correlation coefficients (PCCs) between predicted malignancy scores and the expression of (a) TME-related and (b) tumor-related genes, comparing our teacher model with the baseline SOTA (MCAT). Each point corresponds to an individual gene (gray for Tumor-related gene and yellow for TME-related gene).} \label{fig:PCCs}
\end{figure}

To evaluate the biological relevance of model prediction, we computed Pearson correlation coefficients (PCCs) between predicted malignancy scores and gene expression profiles, focusing on tumor- or TME-related genes. As shown in Fig.~\ref{fig:PCCs}, our teacher model consistently outperforms the SOTA multi-modal baseline (MCAT) in terms of gene-level correlation, suggesting superior alignment with underlying molecular profiles.

Among the most correlated genes, \textit{PTTG1} (PCC = 0.72) and \textit{IFNGR2} (PCC = 0.80) exhibited the highest PCCs within tumor-related and TME-related categories, respectively. This suggests that the model effectively captures molecular signals associated with both intrinsic tumor activity and microenvironmental dynamics.

\subsubsection{WSI-level Interpretability}
\begin{figure}[!t]
\centering
\includegraphics[scale=.295]{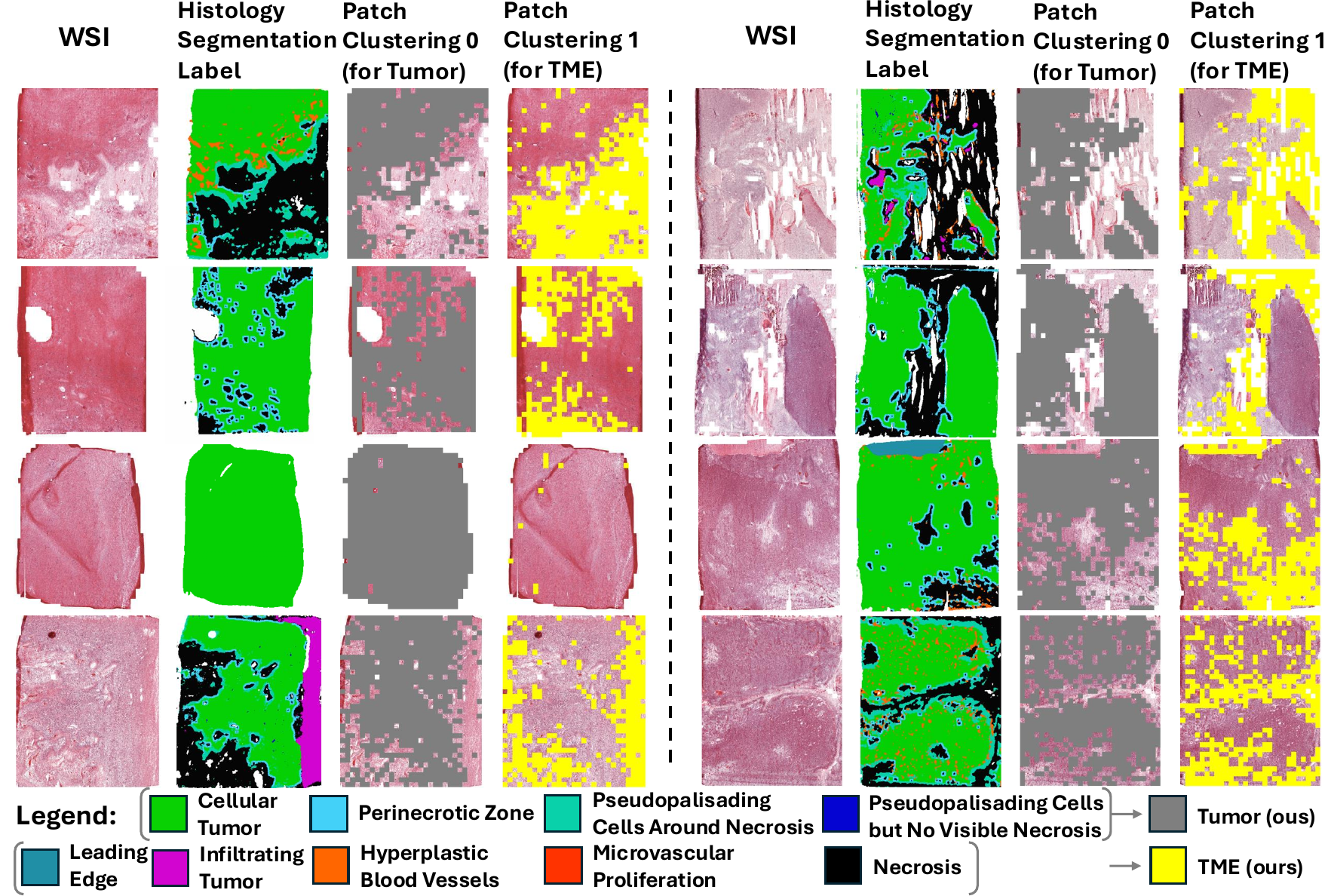}
\caption{Patch clustering on WSIs using distilled student model on IvyGAP dataset. From left to right: raw WSI, histology segmentation label, patch clustering in tumor subspace, patch clustering in TME subspace. Clustering aligns with known histological compartments, demonstrating successful semantic inheritance through cross-modal distillation.}
\label{fig:clustering}
\end{figure}

To evaluate interpretability at the histology level, we visualized the patch clustering outputs from the distilled student model in Stage II (Fig.~\ref{fig:clustering}). Despite operating without transcriptomic input, the model produces tumor- and TME-specific clusters that closely align with expert-annotated regions in the IvyGAP dataset \cite{mohanrajglioblast}. This indicates that the student effectively inherits subspace-specific semantics from the multi-modal teacher. Tumor-related clusters corresponded to expert-labeled regions, such as \textit{Cellular Tumor}, \textit{Perinecrotic Zone}, \textit{Pseudopalisading Cells Around Necrosis}, and \textit{Pseudopalisading Cells but No Visible Necrosis}, all marked by dense tumor cellularity. In contrast, TME-related clusters aligned with regions including the \textit{Leading Edge} (a few tumor cells per 100 normal cells), \textit{Infiltrating Tumor} (10-20 tumor cells per 100 normal cells), \textit{Hyperplastic Blood Vessels}, \textit{Microvascular Proliferation}, and \textit{Necrosis}, reflecting key components of TME \cite{li2019decoding}.
Quantitatively, tumor clusters achieve an average Dice coefficient of 0.52 and a Recall of 0.71, indicating strong sensitivity to malignant regions. For TME clusters, despite greater spatial heterogeneity, the model achieves an average classification accuracy of 0.60. These results suggest that our distillation enables the student model to organize histological patches into biologically meaningful subspaces, even without transcriptomic input.

\section{Conclusions}
This study introduces a biologically inspired, two-stage multi-modal learning framework for cancer characterization that integrates histology and transcriptomics while enabling robust WSI-only inference.
To address the key challenges in multi-modal modeling, integration, and applicability, in Stage I, we first introduce a disentangled learning strategy that decomposes multi-modal features into tumor and TME subspaces through the DMSF module, and coordinates subspace optimization using a CGC strategy. Meanwhile, multi-scale integration is enhanced by an IGC strategy. Stage II facilitates WSI-based inference by combining the ITA module and SKD strategy. 
Our results show that these designs contributed to consistently superior performance across diagnosis, grading, and survival predictions over other SOTA methods under uni-modal, missing-modality, and multi-modal settings. Notably, our distilled model also achieves competitive performance using WSI alone, highlighting its translational potential where transcriptomics are unavailable. External evaluation on unseen data further confirms the generalizability of our teacher model, underscoring the robust learned representations for translation. 

We have several limitations. 
First, our multi-modal training relied on paired modalities. Future work may leverage generative modeling to include both paired and unpaired data for training. 
Second, while our model performs well on three glioma tasks, broader validation on pan-cancer datasets is needed to assess its scalability. 
Finally, while the proposed modules are motivated by common subspace characteristics, future work may incorporate more fine-grained subspaces to enable deeper biological alignments and explanations.  

\appendices




\bibliographystyle{IEEEtran}
\bibliography{IEEEabrv,refs}

\end{document}